\documentclass[journal=jcisd8,manuscript=article]{achemso}
\setkeys{acs}{articletitle = true}

\usepackage[version=3]{mhchem} % Formula subscripts using \ce{}
\usepackage{amsfonts}
\usepackage{makecell}
\usepackage{graphicx}
\usepackage{caption}
\usepackage{hyperref}
\usepackage{placeins}
\usepackage{longtable}
\usepackage{url}
\usepackage{enumerate}
\usepackage[export]{adjustbox}
\usepackage{float}
\usepackage{graphicx}
\usepackage{subfigure}
\usepackage{framed}
\usepackage{amsthm}
\usepackage{appendix}
\usepackage{amssymb}
\usepackage{amsfonts}
\usepackage{enumerate}
\usepackage{algorithm}
\usepackage{algorithmic}
\usepackage{amsmath}
\usepackage[utf8]{inputenc}
\usepackage{amsthm,amssymb}
\usepackage{graphicx}
\usepackage{indentfirst}
\usepackage{dsfont}
\usepackage{placeins}
\usepackage{multirow}
\usepackage{seqsplit}

\author{Wenhao Gao}
\affiliation[MIT]{Department of Chemical Engineering, MIT, Cambridge, MA}
\altaffiliation{Department of Chemical and Biomolecular Engineering, Johns Hopkins University, Baltimore, MD}

\author{Connor W. Coley}
\email{ccoley@mit.edu}
\affiliation[MIT]{Department of Chemical Engineering, MIT, Cambridge, MA}
\alsoaffiliation{Broad Institute of Harvard and MIT, Cambridge, MA}

\title{The Synthesizability of Molecules Proposed by Generative Models}

\keywords{Synthetic feasibility, Molecular optimization, Retrosynthesis planning, Cheminformatics}

\begin{document}
%%%%%%%%%%%%%%%%%%%%%%%%%%%%%%%%%%%%%%%%%%%%%%%%%%%%%%%%%%%%%%%%%%%%%
%% The manuscript does not need to include \maketitle, which is
%% executed automatically.  The document should begin with an
%% abstract, if appropriate.  If one is given and should not be, the
%% contents will be gobbled.
%%%%%%%%%%%%%%%%%%%%%%%%%%%%%%%%%%%%%%%%%%%%%%%%%%%%%%%%%%%%%%%%%%%%%
\begin{abstract}

The discovery of functional molecules is an expensive and time-consuming process, exemplified by the rising costs of small molecule therapeutic discovery. One class of techniques of growing interest for early-stage drug discovery is \emph{de novo} molecular generation and optimization, catalyzed by the development of new deep learning approaches.\cite{Sanchez-Lengeling2018} These techniques can suggest novel molecular structures intended to maximize a multi-objective function, e.g., suitability as a therapeutic against a particular target,\cite{zhavoronkov2019deep} without relying on brute-force exploration of a chemical space.\cite{lyu_ultra-large_2019} However, the utility of these approaches is stymied by ignorance of synthesizability. To highlight the severity of this issue, we use a data-driven computer-aided synthesis planning program\cite{coley2019robotic} to quantify how often molecules proposed by state-of-the-art generative models cannot be readily synthesized. Our analysis demonstrates that  there are several tasks for which these models generate unrealistic molecular structures despite performing well on popular quantitative benchmarks. Synthetic complexity heuristics can successfully bias generation toward synthetically-tractable chemical space, although doing so necessarily detracts from the primary objective. This analysis suggests that to improve the utility of these models in real discovery workflows, new algorithm development is warranted.

\end{abstract}

\section{Introduction}

%%%%%%%%%%%%%%%%%%%%%%%%%%%%%%%%%%%%%%%%%%%%%%%%%%%%%%%%%%%%%%%%%%%%%
% Defect of HTVS

Molecular design is one of the most fundamental challenges in chemical science and engineering. This task is to identify one or more molecules with a specific set of properties of interest, such as binding affinity and drug-likeness for drug design. High-throughput virtual screening (VS) is one widely used strategy to coarsely optimize a molecular structure using a discretized subspace of the whole chemical space.\cite{Walters2019} In VS, we evaluate enumerated candidate molecules in terms of their predicted properties of interest and ranked for follow-up experimental validation. %This strategy has led to the successful identification of hits in combination with docking models.\cite{konze2019reaction}  
However, because we rarely know \emph{a priori} where the ideal molecule will be within the massive design space of chemical space, there is a trend toward using exceedingly large virtual libraries to increase the likelihood that we will find promising candidates. Modern virtual libraries may comprise hundreds of millions or billions of candidate molecules,\cite{lyu_ultra-large_2019} often generated through combinatorial enumeration of commercially-available building block compounds. Even billions of compounds, however, represent a tiny fraction of theoretically-possible, pharmacologically-relevant small molecules, often cited as exceeding $10^{60}$ structures.\cite{virshup2013stochastic} Brute-force virtual screening screening over a chemical space of this size is clearly computationally intractable.

%%%%%%%%%%%%%%%%%%%%%%%%%%%%%%%%%%%%%%%%%%%%%%%%%%%%%%%%%%%%%%%%%%%%%
% de novo design Generative methods: why long-term preferable 

Recent developments in computer aided drug design (CADD) techniques, especially in \emph{de novo} molecular generation and optimization methods, raise the hope of removing this bottleneck.\cite{Sanchez-Lengeling2018} Generative algorithms are a class of methods that propose molecular structures in a manner that can be tailored toward a specific objective. There is a long history of generative models in chemistry, many based on genetic algorithms \cite{lameijer2005evolutionary} and the iterative construction of molecules from molecular fragments.\cite{lewis_novo_1996}  In the past decade, following on the advent of Variational Auto-Encoders (VAEs)\cite{kingma2013auto} and Generative Adversarial Networks (GANs)\cite{goodfellow2014generative}, there has been a flood of new deep learning (DL) methods for this task.\cite{elton_deep_2019} Many of these methods learn a mapping from a continuous lower-dimensional real number space to a discrete chemical space. Jointly trained with a structure-property regression, one can obtain novel chemical structures conditioned on desired properties. More usefully, combining generative models with Bayesian optimization (BO), or directly using a heuristic optimization algorithm (e.g., a genetic algorithm (GA) or tree search (TS)), we can bias candidate generation toward the functionality we desire. Deep generative models are trained on a finite set of molecules to learn an underlying distribution of chemical space, where interpolation and extrapolation produce novel chemical structures.  Enumerating every candidate molecule is thus unnecessary, and applying these models requires linear computational cost to generate multiple molecular structures once trained. Further, the generative algorithms can explore chemical space beyond the limited beginning pool and provide novel chemical structures with preferential intellectual property (IP) positions, whereas molecules in VS are often pre-existing. In recent years, generative models have been applied to various chemical discovery problems and have shown promise as a useful tool for the problem of molecular design.\cite{zhavoronkov2019deep} 

%%%%%%%%%%%%%%%%%%%%%%%%%%%%%%%%%%%%%%%%%%%%%%%%%%%%%%%%%%%%%%%%%%%%%
% de novo design Generative methods: limitation, synthesizability

However, a practical problem that obstructs the usefulness of generative algorithms is that proposed molecular structures may be challenging or infeasible to synthesize. In any realistic discovery scenario, we will need to validate whether a proposed molecule has the property profile we expect; even if our computational models are infallible, we will need to manufacture the molecule in order to apply it (e.g., as a therapeutic, as a catalyst, as a component of a device). Libraries for virtual screening can be constructed from commercially-available databases. They are often enumerated using well-characterized reaction templates to \emph{try} to ensure that enumerated molecules are readily synthesizable. \citeauthor{lyu_ultra-large_2019} report an 86\% successfully synthesis rate among 51 top-ranking molecules from a library comprising 99 million structures, consistent with the claims of many chemical vendors.\cite{lyu_ultra-large_2019}  

The situation is quite different in \emph{de novo} molecular design, especially when using deep generative methods. We expect (and want) these models to explore molecular structures beyond the ones they have been trained on, so they may propose nonsensical structures that are unreasonable for pharmaceutical purposes. There have been few studies explicitly examining this problem, but some anecdotal evidence suggests that compounds are not easily synthesizable---many structures reported in papers indeed appear absurd. \citeauthor{bjerrum_molecular_2017} examined 21 molecules proposed by their recurrent neural network (RNN) model with Wiley's ChemPlanner and found a number of possible selectivity issues in the proposed syntheses, indicating synthetic difficulty.\cite{bjerrum_molecular_2017} \citeauthor{sumita_hunting_2018} filter generated molecules by requiring that they be previously reported with at least one synthetic route in SciFinder, which removes these models' ability to propose novel chemical structures.\cite{sumita_hunting_2018} \citeauthor{zhavoronkov2019deep} select only 6 molecules from 40 candidates structures based on synthetic accessibility, even after filtering an initial list of 30,000 structures generated by a deep learning model.\cite{zhavoronkov2019deep}

%This \textit{post hoc} filtering requires ......

% Besides, because of the high false-positive rate, this synthesizability analysis would also help HTVS.

%%%%%%%%%%%%%%%%%%%%%%%%%%%%%%%%%%%%%%%%%%%%%%%%%%%%%%%%%%%%%%%%%%%%%
% Synthesizability problem and explicit retrosynthesis planning software, how to improve these models

% Synthesizability is an objective concept without an unambiguous definition. First, synthesizability relates to the experimental difficulty of producing a molecule, and varies by personal ability and material availability. Second, synthesizability is not a smooth function with respect to molecular structure. Two similar structures with a single functional group transposition can require substantially different synthetic routes due to the selectivity of chemical reactions, which makes it challenging to fit a good proxy score. (See Figure\ref{fig:comp}) Furthermore, synthesizability is continually evolving with time due to the discovery of new synthetic strategies and expanding chemical vendor catalogs. 

% Some famous examples include Ertl and Schuffenhauer's SA\_Score metric based on structural motif frequency in the PubChem database, and Coley's SCScore based on reactions in Reaxys data.

Current  procedures for quantifying synthesizability are based on (1) structure complexity and similarity or (2) synthetic pathways. The structure-based approach usually involves constructing a heuristic definition based on domain expertise or chemical substructure diversity\cite{bertz_first_1981,ertl_estimation_2009} or designing a model that can be fit to expert scores\cite{takaoka2003development, sheridan_modeling_2014, baba_wisdom_2018} or reaction data.\cite{li_current_2015, coley2018scscore} This kind of method is widely used due to its ease of implementation and low computational cost. However, two similar structures with a single functional group transposition can require substantially different synthetic routes (e.g., due to the selectivity of chemical reactions or availability of specific building blocks), which makes it challenging to fit a good proxy score (see Figure~\ref{fig:comp} and \ref{fig:path} for one example). The most convincing metric might be a direct scoring from a group of experts on synthetic, medicinal chemistry, which has been used as a ground truth to train models against.\cite{takaoka2003development, sheridan_modeling_2014, baba_wisdom_2018, li_current_2015} To have a group of experts that large enough to reach a non-biased and stable value is labor-intensive, hard to replicate, and not scalable. \cite{lajiness2004assessment}

The second, more nuanced approach to measuring synthesizability is to explicitly plan a synthetic pathway and assess its likelihood of experimental validity. Synthetic pathway-based approaches can incorporate more thorough information about starting materials and chemical reactions, which enable them to overcome the shortcomings of the structure-based analysis. In this approach, a computer-aided synthesis planning (CASP) program \cite{feng2018computational} can be used to perform the retrosynthetic analysis. Using an explicit CASP tool is  capable not only  of capturing the high ``non-linearity'' of  synthesizability with respect to chemical structure, but of recommending actionable synthetic pathways. We see this as a form of interpretability to verify \emph{why} the molecule is believed to be synthesizable, with which building blocks, and in how many steps. Only a handful of studies have used a retrosynthetic planning tool to analyze synthesizability\cite{podolyan_assessing_2010, huang_rasa:_2011, bonnet2012chemical, bjerrum_molecular_2017}. Its practical application in molecular design is not widespread yet. Therefore, here, we analyze synthesizability of compounds proposed through generative algorithms using our open-source computer-aided retrosynthesis analysis tool, ASKCOS.\cite{coley2019robotic}

%However, as mentioned above, a good proxy score is not easy to obtain and, more importantly, synthetic complexity is not the same as structural complexity. 

%While a total synthesis of a steroid with multiple chiral center is challenging, a synthesis starting from a readily available compound like cholesterol might only require a few steps. 

% due to the following reasons. First, many retrosynthetic analysis systems, such as WODCA, LHASA, and SECS, are interactivesystem that require considerable knowledge of organic chemistry. Second, they cannot analyze a compound within a reasonable time-scale. Further, the quality of results of many CASP tools is not very high and chemists don't usually trust them.

We divide our analysis of the synthesizability of molecules generated by \emph{de novo} generative algorithms into evaluations of \textit{distribution learning} and \textit{goal-directed generation} tasks--unoptimized and optimized molecules, respectively. Distribution learning models are meant to interpolate within a chemical space comprised a training set of molecules and generate new molecules with similar properties. Goal-directed generation instead tries to generate new molecules that maximize a black-box scoring function. There are an increasing number of algorithms of these two categories proposed in recent years and a small number of studies that benchmark these algorithms in terms of their ability to generate novel, optimal molecules.\cite{polykovskiy_molecular_2018, brown_guacamol:_2018}

\begin{figure}[h!]
    \centering
    \includegraphics[width=\textwidth]{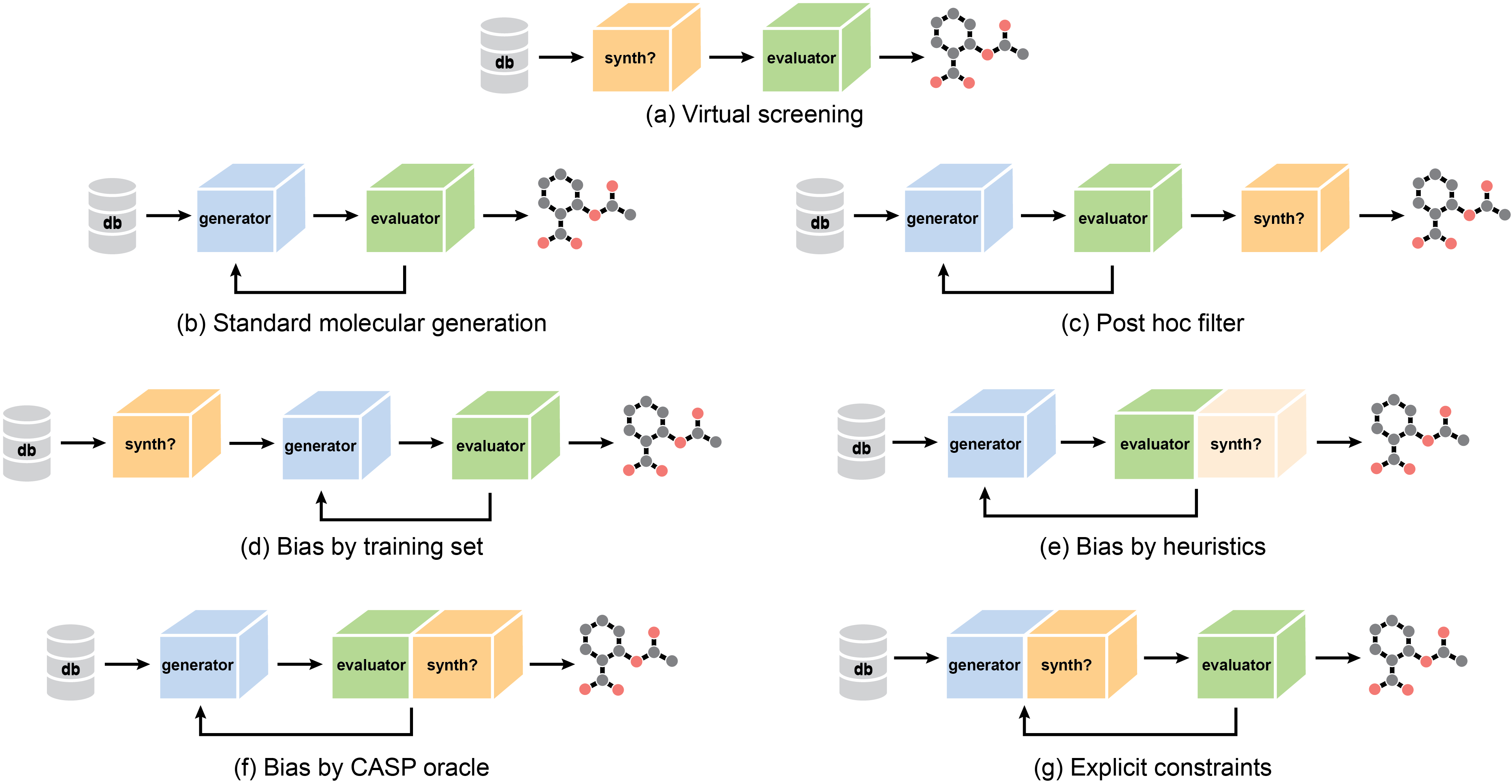}
    \caption{Schematic representation of  approaches to address the challenge of synthesizability in molecular optimization: (a) virtual screening can use a filtered database of candidates to ensure that they are all synthetically accessible; (b) standard molecular generation focuses on evaluation of properties without regard for synthesizability; (c) a \textit{post hoc} filter narrows down proposed candidates as a separate step from generation; (d) biasing by training set aims to improve synthesizability by training generative models on synthetically-accessible compounds; (e) biasing by heuristics uses simple scalar proxies for synthesizability as part of the objective function; (f) biasing by a CASP oracle runs a full retrosynthetic expansion for proposed molecules to modify the reward function in a reinforcement learning setting; and (g) explicit constraints attempt to restrict chemical space to what is accessible using buyable building blocks and known synthetic transformations.}
    \label{fig:schematics}
\end{figure}

We categorize the approaches one might take to ensure that computationally designed molecules are able to be synthesized in Figure~\ref{fig:schematics}. These represent combinations of (i) a database of known or enumerated compounds, (ii) an evaluator, which estimates the properties we are trying to optimize, (iii) a generator function, which can propose new candidate molecules, (iv) a synthesizability oracle that determines whether it is straightforward to synthesize a given molecule, and/or (v) a heuristic synthesizability estimator that provides a computationally-inexpensive scalar measure of synthesizability.  In this study, we focus on  three major approaches to solving the synthesizability problem: \textit{post hoc} filtering (Figure~\ref{fig:schematics}c), imposing \emph{a priori} differences in training sets (Figure~\ref{fig:schematics}d), and heuristic biasing (Figure~\ref{fig:schematics}e). %% For \textit{post hoc} filtering, we analyze the generation result from both distribution learning and goal-directed optimization. We separately train models on ChEMBL (less synthesizable) and MOSES (more synthesizable) and compare the generation result to simulate the approach of  \emph{a priori} biasing by training set. For heuristic biasing, we apply a synthesizability function multiplier, ranging from 0 to 1, to a pre-normalized objective function to enhance the sampling of highly synthesizable chemical space.

%So we first demonstrate the validity of using retrosynthetic analysis to quantify synthesizability, even with this imperfect tool, and analyzed the synthesizability of common chemical data sets. Further, we analyzed current state-of-art molecular generation algorithms and compared several solutions (see Figure \ref{fig:schematics}) to tackle this problem. 

\section{Results}
\subsection{Synthesizability of common databases according to ASKCOS}

% \textbf{Move the paragraph about ASKCOS down here.} \textbf{A detailed description of the search settings used for all evaluations can be found in the Methods section.}

\begin{figure}[th!]
    \centering
    \includegraphics[width=\textwidth]{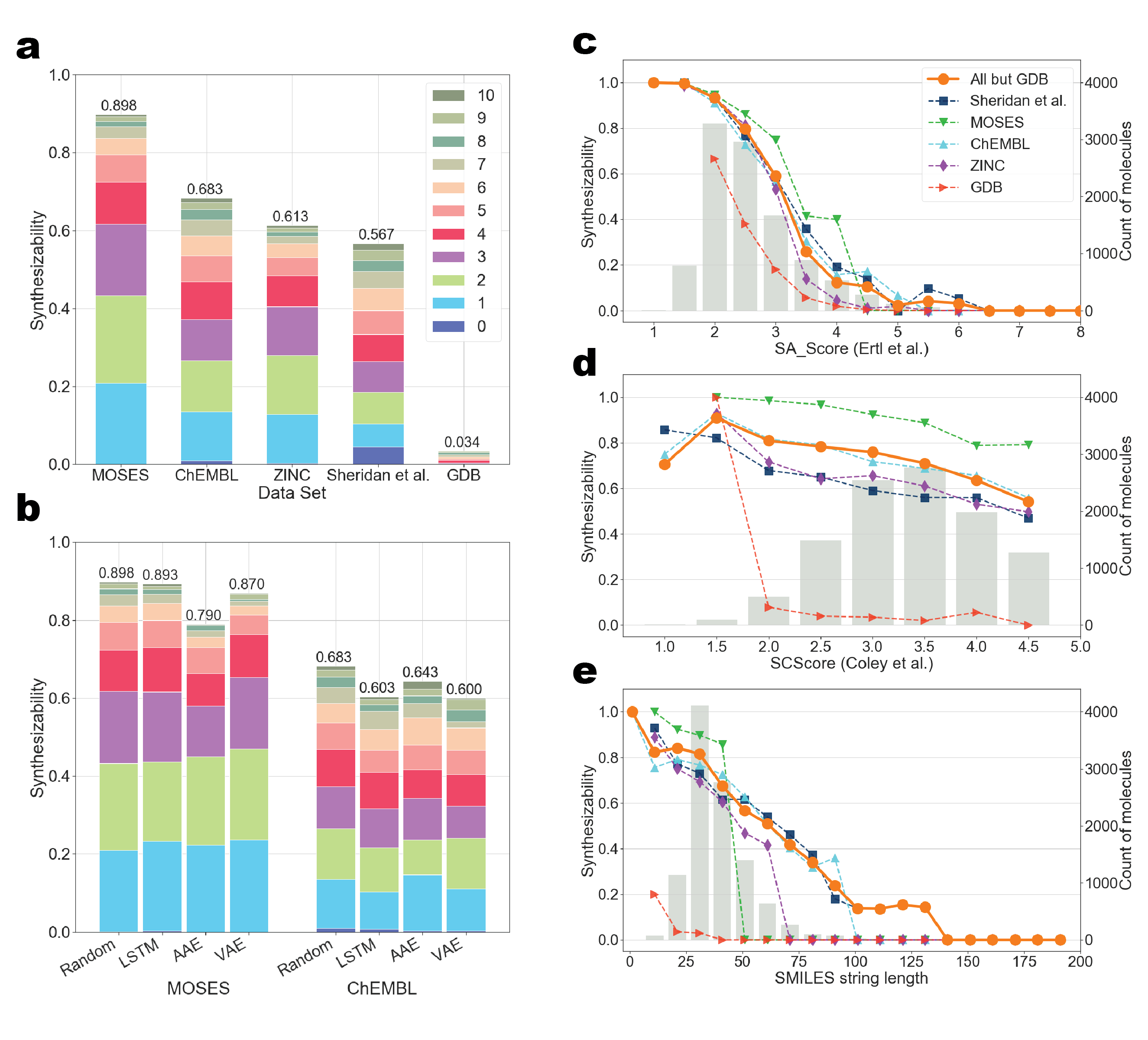}
    \caption{The synthesizability analysis of common data sets, distribution learning, and popular heuristics. (a) the number of synthetic steps required to produce random sampled structures from each data set; (b) the number of synthetic steps required to produce molecules generated by distribution learning algorithms, trained on either MOSES or ChEMBL; (c-e) the fraction of synthesizable compounds from each dataset binned by heuristic score and the number of molecules scored within each bin (excluding GDB).}
    \label{fig:barplot}
\end{figure}

We first validate that the information returned by ASKCOS is usefully correlated with synthesizability by analyzing molecules from several standard compound libraries: MOSES,\cite{polykovskiy_molecular_2018} ChEMBL,\cite{gaulton_chembl:_2012} ZINC,\cite{sterling_zinc_2015} \citeauthor{sheridan_modeling_2014},\cite{sheridan_modeling_2014} and  GDB17\cite{ruddigkeit_enumeration_2012} (see Methods for detailed descriptions of each data set and the settings used for retrosynthetic analysis, including the evaluation of commercial availability of building blocks). Figure~\ref{fig:barplot}a shows the predicted number of synthetic steps required to produce a random set of 3000 molecules from each data set. The MOSES data set has the highest rate of perceived synthesizability at 89.8\%, consistent with its focus on small lead molecules and exclusion of compounds with ``structural alerts''. Its parent set, ZINC, has a lower synthesizability rate of 60.8\%. The ChEMBL data set has a higher rate of 68.3\%; although it contains larger and more complex structures than does ZINC, many have been synthesized previously; among those that cannot be synthesized are natural products that were extracted, not synthesized, and tested for their biological activity. ChEMBL also contains several directly purchasable compounds, second only to \citeauthor{sheridan_modeling_2014}'s data set of 1730 compounds.  Unsurprisingly, the exhaustively enumerated data set, GDB17, has the lowest rate of synthesizability at only 3.5\%. We also find that the predicted  number of reaction steps is correlated with expert-provided scores (Figure~\ref{fig:depth_mc}). From these trends and the high success rate of the MOSES database, we conclude that ASKCOS's retrosynthetic analyses are largely consistent with our expectations of synthesizability and it is appropriate to use its predictions to benchmark the evaluation of molecular generation.

\subsection{Agreement between synthesizability heuristics and ASKCOS}

We next evaluate the agreement between several heuristic synthesizability scores (length of SMILES, SA\_Score\cite{ertl2009estimation}, and SCScore\cite{coley2018scscore}) and the results of ASKCOS. Because retrosynthetic analysis can be time consuming (tens to hundreds of CPU-seconds), we would prefer to bias generation by heuristics rather than by a CASP oracle (cf. Figure~\ref{fig:schematics}). 
Figure~\ref{fig:barplot}c-e show the trend of synthesizability of structures in different range of SA\_Score, SCScore, and SMILES string length. None of them can distinguish the synthesizable and unsynthesizable compounds perfectly, but all exhibit a decreasing trend as the heuristic score increase. The trend is clearest for the SA\_Score, followed by the SMILES length and then the SCScore. This ordering is quantified in Figure~\ref{fig:roc}, which shows the area under the receiver operating characteristic as if these heuristics were being used for binary classification. The AUC values for the three methods in this order are 0.87, 0.69, and 0.61. 
%While the curves of SCScore and SMILES string length are more vague, SA\_Score has a much smoother and more monotonic curve, despite of that SA\_Score is all about structural complexity. Even the least synthesizable data set, GDB, has the similar trend, just constantly lower than the general curve. Structures with SA\_Score in range of 4.5-5.0 only have less than a 10\% probability of being synthetically accessible. 
The slight shoulder around 5.5-6.0 is the contribution from the structurally complex but commercially available compounds, highlighting the difference between synthetic complexity and structural complexity as discussed in ref.~\citenum{coley2018scscore}.

\subsection{Synthesizability of unoptimized generated molecules}

% To assess the synthesizability of molecules generated by \emph{de novo} generative algorithms, we separately tested distribution learning and goal-directed generation to look at unoptimized and optimized molecules. As mentioned above, distribution learning models are meant to interpolate within a chemical space comprised a training set of molecules and generate new molecules with similar properties. Goal-directed generations instead aim to generate new molecules that maximize a black-box scoring function. 

As alluded to above, distribution learning methods are capable of generating ``unoptimized'' molecules that share properties (in aggregate) with the database used for training. Here, we evaluate methods implemented in the MOSES\cite{polykovskiy_molecular_2018} benchmarking set, which cover diverse approaches to the molecular generation problem: a SMILES long short-term memory (LSTM) model, a variational auto-encoder (VAE), and an adversarial auto-encoder (AAE) (see Methods). There are more deep learning approaches for molecular generation and optimization than can be compared here,\cite{elton_deep_2019} so we focus on these top-performing classes of approaches. In this task, we can use \textit{post hoc} filtering or training set biasing by separately training distribution learning models on ChEMBL (less synthesizable) and MOSES (more synthesizable).

Figure~\ref{fig:barplot}b shows the fraction of synthesizable molecules from 300 generated by each distribution learning method trained on the ChEMBL and MOSES. We observe that the fraction of synthesizable molecules are comparable to that of the training set, while no method  \emph{improves} synthesizability relative to its training set. The stark difference between results using MOSES and ChEMBL suggests that  \emph{a priori} biasing by training on a ``more synthesizable'' data set is a viable approach for distribution learning algorithms. There is no one method particularly superior than others. The high fraction of synthesizable results further suggests that \textit{post hoc} filtering is not necessarily a bad approach (i.e., relatively few generated molecules would fail a check for synthesizability). Note these results pertain only to the synthesizability of generated results and do not consider previously evaluated metrics of  novelty, uniqueness and diversity as do \citeauthor{polykovskiy_molecular_2018}'s analyses.\cite{polykovskiy_molecular_2018}  

\subsection{Synthesizability of optimized generated molecules}

\begin{figure}[h!]
    \centering
    \includegraphics[width=\textwidth]{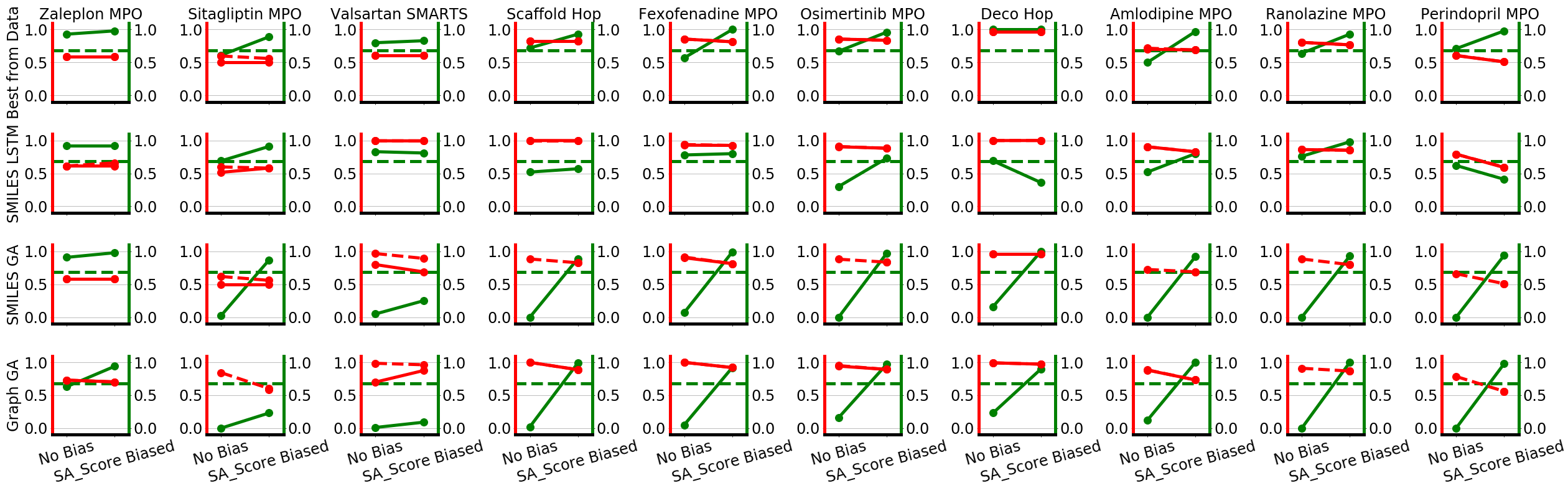}
    \caption{Dependence of goal-directed optimization performance on heuristic biasing by the SA\_Score using ChEMBL as the training database. Each row represents result from one generative method; each column represents one objective function. In each plot, the green solid line represents the change of fraction of synthesizable compounds in the top-100 with the green dashed line as a reference for the synthesizability of the training set (ChEMBL). Red solid lines represent the change in the objective function value of the top \emph{synthesizable} molecule, while the dashed red line represents the change in  objective function value of the top molecule, regardless of its synthesizability. Plots without a solid red line indicates that no synthesizable structure was obtained in the top 100 molecules; dashed red lines may be occluded by solid red lines.}
    \label{fig:goal_change}
\end{figure}

\begin{figure}[h!]
    \centering
    \includegraphics[width=\textwidth]{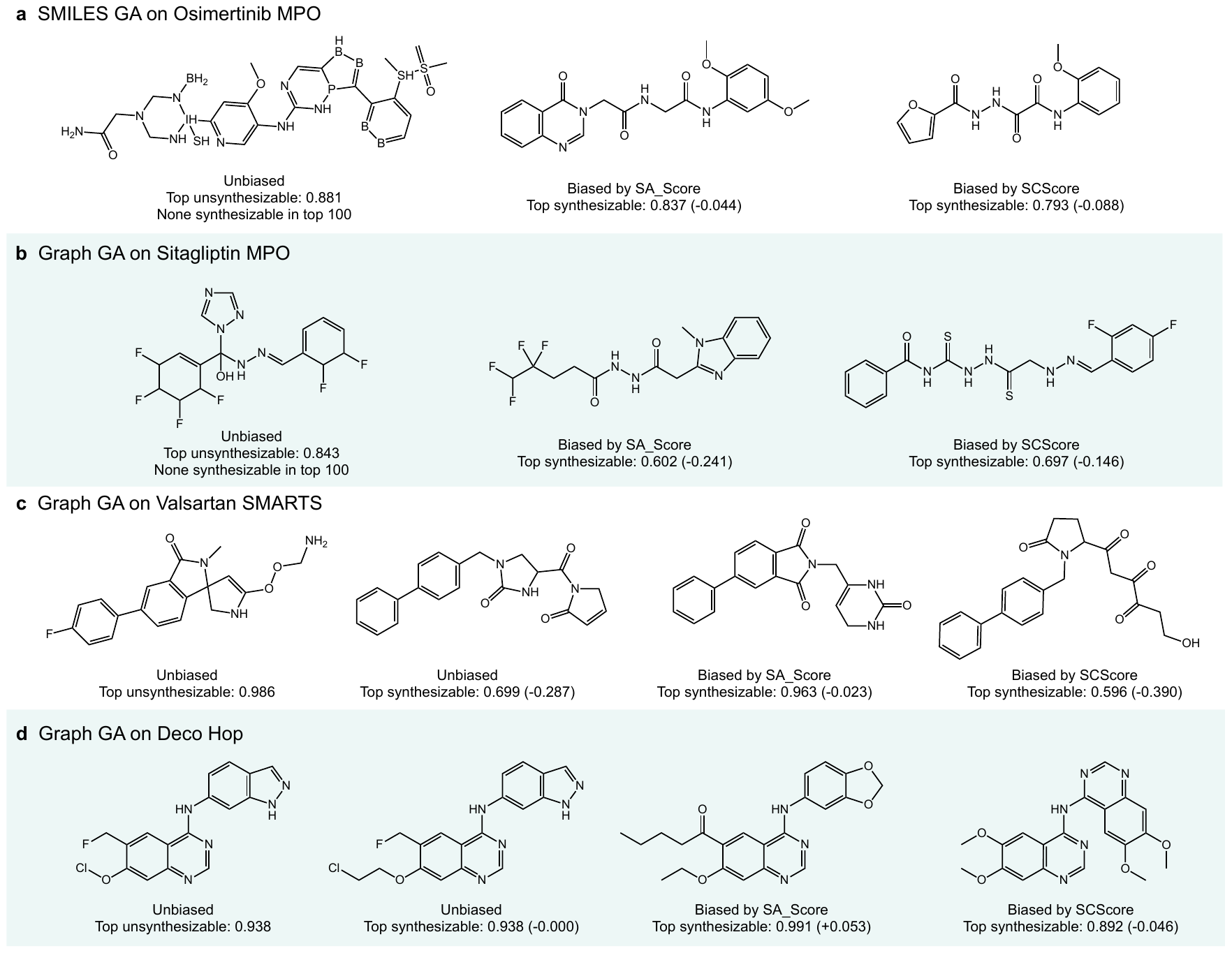}
    \caption{Examples of molecules from goal-directed optimization that were improved by heuristic biasing. Scores shown are the objective function values that have been normalized to the interval [0, 1]. (a,b) cases where no synthesizable compounds were found in the top 100 suggestions without biasing, but at least one was found with either SA\_Score or SCScore biasing. (c,d) cases where the top structure found without biasing was perceived as unsynthesizable and the use of heuristic biasing improved the objective function value of the top \emph{synthesizable} structure.}
    \label{fig:goal_mols}
\end{figure}

% What we did

Our next analyses focus on \emph{goal-directed benchmarks}, which reflect the actual use-case for generative models. Here, we re-evaluate the  methods and objective functions evaluated by \citeauthor{brown_guacamol:_2018}'s  Guacamol\cite{brown_guacamol:_2018} in terms of their synthesizability. As detailed in the Methods, this includes three generative algorithms (SMILES LSTM, SMILES GA, and Graph GA) and 14 multi-property objective functions (MPOs) that convert a molecular structure to a scalar fitness score. As a baseline method, we include a virtual screening approach, ``Best from Data'', where all candidates from either ChEMBL or MOSES are evaluated to identify the top performers. In addition to \textit{post hoc} filtering and training set biasing, we can also bias generation by modifying the objective function with a heuristic synthesizability score.  We multiply the original objective functions (normalized between 0 and 1) with a quantitative synthesizability metric (SA\_Score or SCScore, also normalized between 0 and 1). More details can be found in Methods section.

We evaluate the effects of heuristic biasing both in terms of the synthesizability of suggested molecules and in terms of the primary objective function value. Figure~\ref{fig:goal_change} shows how these metrics change when biasing with SA\_Score, initially trained on ChEMBL (Figure~\ref{fig:plot_change} shows additional results using the SCScore/ChEMBL, SA\_Score/MOSES, and {SCScore/MOSES}). Compared to the synthesizability of starting set (the green dashed lines), we can see the synthesizability varies between different methods and objectives. Indeed, the total fraction of synthesizable compounds in all methods for ``hard'' objectives without biasing is 30.2\% with ChEMBL and 32.7\% with MOSES (see Figure~\ref{fig:hard_synth} for more details), excluding the direct sampling from data set. Compared to distribution learning, the goal-directed generation methods are less sensitive to the starting set of molecular compounds. For several tasks (Figure~\ref{fig:hard_synth}), very few or no compounds in the top 100 are synthesizable in the absence of heuristic biasing, particularly when using the genetic algorithms, illustrating the risk of relying on a \textit{post hoc} filtering strategy. Examples in Figure~\ref{fig:goal_mols}ab illustrate cases where no molecule in the top 100 is synthesizable and heuristic biasing is required to generate even a single feasible candidate. The compounds remaining after filtering for synthesizability, if any, may have low objective function values.

Most cases in Figure~\ref{fig:goal_change} show that the synthesizability of the top 100 compounds after biasing is quite high, often exceeding the rate for ChEMBL. Generally speaking, the SA\_Score performs better than SCScore (Figure~\ref{fig:plot_change}): the overall synthesizability for hard objectives was improved from 30.2\% to 80.2\% or 55.4\% when biasing by SA\_Score or SCScore, respectively, originally trained on ChEMBL (Table~\ref{tbl:heuristic_biasing}). This successful result validates the approach shown in Figure~\ref{fig:schematics}e, but the increased synthesizability comes at the expense of the objective function value of the top candidate. For some tasks decreases by over 0.2--a significant difference for these benchmark tasks. However, we note that the value of an \textit{in silico} objective function is completely inconsequential if the molecule cannot be made and experimentally tested.

\begin{table}[h!]
\caption{Fraction of synthesizable compounds in the top-100 candidates across all goal-directed optimization tasks and all methods, demonstrating successful heuristic biasing.}
\label{tbl:heuristic_biasing}
\begin{tabular}{llrrr}
\hline
                               Training database    &       Task difficulty    & Unbiased & Biased by SA\_Score & Biased by SCScore \\ \hline
\multirow{2}{*}{ChEMBL} & trivial & 60.1\%   & 91.0\%              & 77.9\%            \\
                                   & hard    & 30.2\%   & 80.2\%              & 55.4\%            \\ \hline
\multirow{2}{*}{MOSES}  & trivial & 63.5\%   & 92.2\%              & 78.8\%            \\
                                   & hard    & 32.7\%   & 77.2\%              & 58.0\%            \\ \hline
\end{tabular}
\end{table}

A fairer comparison can be made between the objective function values of the top \emph{synthesizable} candidates, i.e., after \textit{post hoc} filtering.   Figure~\ref{fig:goal_mols}cd shows two examples where the objective of the top-1 candidate decreased, but the value of the top-1 synthesizable candidate increased. That this is observed in some cases (also see Figure~\ref{fig:synth2improve}) suggests a practical workflow for molecular optimization: if only a few synthesizable candidates (1-10) are desired,  first optimize without biasing and  filter  unsynthesizable suggestions; if the top synthesizable candidates are  worse than the top unsynthesizable candidates,  repeat the optimization while biasing with the SA\_score.

\section{Discussion of other approaches}

% A separate discussion section where we talk about methods to embed synthesizability more directly? We can also talk about biasing with a CASP oracle to complete our discussion of Figure 1

% What are these final two techniques in Figure 1?  Has anyone made some progress on this front to-date?

% How would we propose to approach them? What is the limitation given the current state of the art? What are the specific challenges that need be addressed?

As described in Figure~\ref{fig:schematics}, there are more ways to improve synthesizability of \emph{de novo} molecular generation algorithms. One promising approach is to bias the generation using a full CASP tool to evaluate synthesizability, instead of a proxy score (Figure~\ref{fig:schematics}f). The advantages are already described above; the disadvantage is the computational expense. %The  \citeauthor{huang_rasa:_2011} attempted to measure synthesizability based on rapid retrosynthesis by reducing the expansion of synthesis tree thus limiting the combinatorial explosion, but the average time for analyzing one compound is still 31 CPU seconds. \cite{huang_rasa:_2011} 
While ASKCOS finds pathways in a few seconds for some molecules, we spend up to one minute evaluating each molecule to reduce the number of false negatives. 

Benchmarking for molecular optimization, in addition to neglecting synthesizability, has largely neglected the number of objective function calls and computational expense. When using genetic algorithms for molecular optimization, we would first select high scoring synthesizable compounds as the initial set to propagate from a pool of up to millions of structures ($\sim 10^{6}$) and then score, at each of hundreds of iterations ($\sim 10^3$),  hundreds of child compounds ($\sim 10^{3}$). In total, we would require millions or at least hundreds of thousands of calls to the CASP oracle. Reinforcement-learning-based optimization methods that outperform Bayesian optimization when using VAEs require one oracle call per iteration, but require hundreds of thousands or millions of iterations (e.g., MolDQN reports the use of 200k function calls\cite{zhou2019optimization}). %One reason for the sample inefficiency of reinforcement-learning-based optimization methods is the models are spending too much repeating efforts to ``learn'' the normality of molecules. \citeauthor{you2018graph} attempted to combine adversarial training to incorporating prior knowledge specified by a dataset of example molecules. \cite{you2018graph} 
One study by \citeauthor{korovina2019chembo}, who propose a method described in the next paragraph, highlight several existing methods that all require $\ge 5$ thousand evaluations for a single task compared to their 100. Based on the machine learning community's broader interest in improving the sample efficiency of reinforcement learning algorithms\cite{ortega2019meta} (thus fewer times calling the oracle) and CASP tools becoming faster, the use of an explicit retrosynthetic planner \emph{during} optimization may become a computationally viable strategy.

The final approach (Figure~\ref{fig:schematics}g) is to embed synthesizability constraints in the generation algorithm itself, i.e., constrain the search space to molecules that can be produced from available building blocks. As early as 2003,  \citeauthor{vinkers_synopsis:_2003} describe the iterative optimization of molecular structure by selecting building blocks to react with a growing molecular structure\cite{vinkers_synopsis:_2003}. More recently, \citeauthor{bradshaw2019model}\cite{bradshaw2019model} propose a model called MoleculeChef that generates a bag of reactants and uses a forward reaction prediction software to obtain the final products. \citeauthor{korovina2019chembo}'s ChemBO  similarly treats molecular generation as a random walk on a directed (synthetic) graph where each node is a molecule, and the parents of this node are the reagents that produce the child molecule when combined.\cite{korovina2019chembo} These techniques are philosophically aligned with our use of retrosynthetic analysis to evaluate synthesizability--both try to use our collective knowledge of chemical reactivity to dictate what reactions are possible--but operate in the \emph{forward} synthetic direction. %This class of approaches rely on frameworks of reactions and synthetic transformations instead of vertices and edges in molecular graphs or tokens in SMILES strings, so they cannot be directly combined with popular generative approaches and warrant new algorithm development.  
This makes them subject to the same caveats that any CASP tool is subject to: their validity is entirely dependent on the accuracy of their forward reaction prediction engine, which can  use either hand-coded rules or algorithmically-inferred rules. The greater the number of synthetic steps we allow, the lower the chances that each reaction will proceed as predicted. As this is essentially how virtual libraries are constructed, we would expect a similar rate of success (anecdotally, ~85\% successful delivery of compounds from a library enumerated with a single synthetic step). Nevertheless, as the search space is directly constrained by these rules, they may enable a more efficient exploration of chemical space. We expect such algorithms  to rapidly grow in popularity as the accuracy of reaction prediction tools improves.\cite{coley_graph-convolutional_2019, schwaller_molecular_2018}

\section{Conclusion}

%%%%%%%%%%%%%%%%%%%%%%%%%%%%%%%%%%%%%%%%%%%%%%%%%%%%%%%%%%%%%%%%%%%%%
% Conclude results

In this paper, we describe an analysis of the synthesizability of \emph{de novo} generative algorithms. We first examined  common chemical compound libraries and used ASKCOS to evaluate their synthesizability. We next evaluated molecules proposed by distribution learning and goal-directed generation methods, with and without biasing by heuristic synthesizability metrics. Distribution learning methods, provided they can learn the chemical distribution of the training set well, seem to generate molecules that are synthesizable with a similar frequency to their training set. Goal-directed generation methods have a significant risk of proposing unsynthesizable structures as their top suggestions, particularly using the SMILES GA or Graph GA methods, but occasionally there may be enough high-performing, synthesizable molecules in the top 100 that \textit{post hoc} filtering (Figure~\ref{fig:schematics}c) is a viable strategy. In other cases, the proposed molecules are so absurd that one immediately recognizes why benchmarking these methods solely in terms of their objective function value is insufficient (e.g., Figure~\ref{fig:nan2synth_1} and \ref{fig:nan2synth_2}). Biasing generation by training set synthesizability (Figure~\ref{fig:schematics}d) works for distribution learning, but does not have a noticeable effect on goal-directed optimization tasks. For some tasks, modifying the objective function with the SA\_Score leads to candidates that outperform those obtained through \textit{post hoc} filtering (Figure~\ref{fig:goal_mols}cd and Figure~\ref{fig:synth2improve}). This heuristic biasing (Figure~\ref{fig:schematics}e) almost always improves the synthesizability of generated candidates, but necessarily detracts from the main objective function. % However, we should note this solution is temporal and don't necessarily solve the problem of synthesizability in goal-directed generation.

% (1) distribution learning methods generate molecules that are pretty synthesizable, since the source datasets are pretty synthesizable; 

% (2) goal-directed optimization sometimes generates synthesizable molecules, but some tasks make them to very weird things; 

% (3) using SA\_Score as a biasing heuristic does an okay job recovering some synthesizability but at the expense of the objective function; 

% (4) if you were to compare performance of filtering (a) unbiased and (b) SA\_Score-biased molecules in terms of synthesizability so that the objective was evaluated as the objective function value for the top-k synthesizable molecules, the result is

%%%%%%%%%%%%%%%%%%%%%%%%%%%%%%%%%%%%%%%%%%%%%%%%%%%%%%%%%%%%%%%%%%%%%
% Other biasing methods

% There are two additional approaches in the bottom row of Figure \ref{fig:schematics} that we have yet to discuss. The first is to bias molecular generation by a CASP oracle in a reinforcment learning setting. Several generative models are designed to use a black box oracle function to guide their optimization\needcite \cite{guimaraes_objective-reinforced_2017, zhou_optimization_2018}

%We expect more works in this direction would come out and solve the problem of synthesizability. 

We acknowledge that the identification of a synthetic pathway by ASKCOS is not a necessary or sufficient condition for synthesizability, nor would the generation of molecular candidates through forward synthesis prediction be a guarantee that those reactions would work experimentally. CASP tools for retrosynthesis and forward synthesis are imperfect. They do not capture our entire knowledge of chemical reactivity and may occasionally produce overly optimistic suggestions (e.g., with respect to selectivity). Further, the ability of CASP programs to find pathways is  sensitive to the precise database of chemicals considered buyable and the settings one chooses for the retrosynthetic expansion. Even with an imperfect CASP tool like ASKCOS, however, we  can obtain a meaningful analysis of synthesizability of generated molecules.

Generative models have a tremendous potential to accelerate molecular discovery. As we improve their ability to propose synthesizable molecules--whether by improving CASP tools for \textit{post hoc} filtering, developing new heuristics for synthesizability, efficiently sampling a CASP oracle to bias generation with reinforcement learning, or designing new algorithms explicitly constrained by predictions of chemical reactivity--their utility and relevance to practical discovery projects will only increase.

%For future directions, besides further improvement of CASP tools, we expect better ways to obtain a reliable synthesizability measurement quickly. Directly learn the result of a CASP tool with an active querying strategy might be a good attemp. It is also a promising approach to explicitly constrain the search space to molecules that can be produced from available building blocks in the generation algorithm. We expect more works in this direction would come out and solve the problem of synthesizability. 

\section{Methods}

\subsection{ASKCOS}

ASKCOS is an open-source software framework that integrates efforts to generalize known chemistry to new substrates by learning to apply retrosynthetic transformations, to identify suitable reaction conditions, and to evaluate whether reactions are likely to be successful when attempted experimentally.\cite{coley2019robotic, coley_askcos_nodate} Data-driven models within ASKCOS are trained on millions of reactions from the U.S. Patent and Trademark Office (USPTO) and Reaxys databases. The core retrosynthetic capabilities rely on the recursive application of algorithmically-extracted reaction templates encoded as SMARTS patterns.  Expansion is parallelized using an upper confidence bound tree search as detailed in the original publication. Importantly,  ASKCOS has both programmatic and graphical interfaces to enable thousands of compounds to be processed without human intervention. The program makes extensive use of RDKit.\cite{Landrum2016RDKit2016_09_4}

While the program offers flexible stopping criteria, we require starting materials to be commercially available according to a 2018 database of molecules from eMolecules or Sigma Aldrich with prices no greater than \$100 per gram; the full list is available in the ASKCOS codebase. This is a very strict price limit in the context of drug discovery, so it warrants two additional comments. First, one could consider most molecules to be ``commercially available'', in that some supplier or contract research organization will agree to produce them at some cost given sufficient lead time. Second, it is straightforward to modify the database of molecules considered commercially available depending on each user's price tolerance and available chemical inventory. 

To determine whether a molecule is ``synthesizable'', we run a retrosynthetic expansion using ASKCOS with the following expansion settings: the maximum search depth--longest linear sequence--is 9, the maximum branching ratio--number of unique precursors to consider at each disconnection--is 25, the maximum wall time of expansion is 60 seconds, the maximum cumulative probability for target is 0.999, the maximum number of templates to apply is 1000, the maximum price for starting materials is \$100/g as described above, the minimum plausibility of reactions--evaluated by a binary classifier as a ``sanity check''--is 0.1. We terminate the search as soon as a pathway is found, rather than continuing to search for a more optimal (e.g., shorter, cheaper) pathway. All retrosynthetic analyses  were carried out in an ASKCOS server on a debian virtual machine running on Google Cloud with 8 cores,  52 GB memory, and no other background tasks.

\subsection{Compound databases}

\begin{itemize}

\item MOSES\cite{polykovskiy_molecular_2018} is an open database included in the MOSES benchmarking platform that evaluates distribution learning algorithms for drug discovery. The database of 1.94 million structures represents a subset of the 4.6 million in the ZINC Clean Leads collection with molar masses of 250-350 g/mol, fewer than 8 rotatable bonds, and a maximum XLogP of 3.5. \citeauthor{polykovskiy_molecular_2018} filtered out molecules containing charged atoms, atoms besides C, N, S, O, F, Cl, Br, and H,  cycles longer than 8 atoms, and molecules containing ``structural alerts'' from medicinal chemistry filters and PAINS filters.

\item ChEMBL\cite{gaulton_chembl:_2012} is a regularly-updated, open access database containing a large number of biologically-relevant compounds and associated assays (e.g., binding and ADMET). In our experiments, we use ChEMBL release 24, which contains 15.2 million activity measurements for 1.8 million compounds.

\item ZINC\cite{irwin2012zinc} is an open database of commercially-available (not in-stock) compounds for virtual screening. ZINC contains over 230 million purchasable compounds in ready-to-dock, 3D formats. We sampled molecules from ZINC-250k, which is a widely used subset of ZINC12\cite{irwin2012zinc} from \citeauthor{gomez2018automatic}.\cite{gomez2018automatic}

\item \citeauthor{sheridan_modeling_2014}\cite{sheridan_modeling_2014} refers to a set of 1730 unique and parseable compounds taken from the 2575 unique molecules released by Merck in their paper exploring a crowdsourced definition of molecular complexity. These molecules were drawn from various public and Merck-internal sources as described in the original publication. 

\item GDB17\cite{ruddigkeit_enumeration_2012} is an open database containing 166.4 billion enumerated molecules with up to 17 heavy atoms of C, N, O, S and halogens. The enumeration started from mathematical graphs to form skeletons, aiming to cover size ranges containing many drugs and typical for lead compounds. We are sampling from its ``Lead-like Set'' of 800 thousand compounds with molar masses of 100-350 g/mol,  CLogP of 1-3, and without 3- or 4-membered rings.

% ChEMBL contains only molecules that have been synthesized and tested against biological targets. ZINC is constructed aiming to cover all drug leads and thus concentrate on smaller molecules, and at least a part of them have not been made yet. MOSES is a filtered version of ZINC for benchmarking distribution learning. GDB17 is an exhaustively enumerated dataset of C,H,N,O,S,halogen-containing species up to 17 atoms, the vast majority of which have not been synthesized and many of which include complex annulated rings systems. We sampled and analyzed 3000 molecules with the same properties distribution from each dataset, and to compare with expert scoring, we also analyzed 1730 molecules from the dataset used in Sheridan's work. (cite)
    
\end{itemize}

\subsection{Molecular generation algorithms}

\begin{itemize} 

\item \textit{Random sampler} is a baseline approach to molecular generation and optimization that randomly samples molecules (with replacement) from a ``training set'' of known compounds.

\item \textit{Best from data} represents the virtual screening approach to molecular optimization, where  all molecules from a ``training set'' of known compounds are evaluated to identify the ones with the highest scores.

\item \textit{LSTM}\cite{segler2017generating} refers to a Long-Short Term Memory\cite{hochreiter1997long} neural network that is widely used in natural language processing. The model is trained in an auto-regressive way to predict the next character of a simplified molecular‑input line‑entry (SMILES) string. It can be iteratively fine-tuned to optimize molecules toward a specific objective using a hill-climbing algorithm. We evaluated the implementation from ref.~\citenum{brown_guacamol:_2018}.

\item \textit{VAE}\cite{gomez2018automatic} refers to a variational autoencoder architecture that learns to construct a bidirectional mapping between SMILES represented chemical space and a finite-dimensional continuous latent space. The architecture is devised to learn a probabilistic generative model as well as its posterior, respectively known as decoder and encoder. The two parts are trained simultaneously by maximizing the evidence lower bound (ELBO) of the marginal likelihood, $ELBO(\phi, \theta) = \mathds{E}_{q_{\phi}(z|x)} [\log p_{\theta}(x|z)] - KL(q_{\phi}(z|x) || p(z))$, where $\phi$ and $\theta$ are differential parameters and KL is the Kullback–Leibler (KL) divergence. We evaluated the implementation from ref.~\citenum{polykovskiy_molecular_2018}.

\item \textit{AAE}\cite{polykovskiy2018entangled} is another approach to train a SMILES-based encoder-decoder architecture. Instead of KL regularization, AAE is trained with an adversarial learning regularization that matches the posterior distribution to a prior distribution. We evaluated the implementation from ref.~\citenum{polykovskiy_molecular_2018}.

\item \textit{SMILES GA}\cite{yoshikawa2018population} is a population-based grammar evolution algorithm. We evaluated \citeauthor{yoshikawa2018population}'s model that adopted a ``chromosome'' with context-free grammar of SMILES string so that crossover and mutation happens at the level of SMILES tokens. Each ``chromosome'' can be decoded to a SMILES string and checked validity using . We evaluated the implementation from ref.~\citenum{brown_guacamol:_2018}.

\item \textit{Graph GA}\cite{jensen2019graph} is another genetic algorithm that represents molecules as graphs, rather than relying on SMILES strings. The crossovers and mutations are performed by altering a molecular graph directly, i.e., exchanging substructures and hand-written substitution rules for mutation. We evaluated the implementation from ref.~\citenum{brown_guacamol:_2018}.

\end{itemize}

% We separately tested distribution learning and goal-directed generation to look at unoptimized and optimized molecules. Distribution learning models are meant to interpolate within a chemical space defined by a training set of molecules and generate new molecules with similar properties.  Goal-directed optimizations instead aim to generate new molecules that maximize a black-box scoring function. We adopted distribution learning methods from MOSES\cite{polykovskiy_molecular_2018} and goal-directed learning methods from Guacamol\cite{brown_guacamol:_2018}, which cover diverse approaches to the molecular generation problem (\#\#\#, \#, and \#\#\#, and \#\#\#). Details of these algorithms can be found in these benchmarking papers. There are more deep learning approaches for molecular generation and optimization than can be compared here,\cite{elton_deep_2019} so we focus on these top-performing classes of approaches. 
\subsection{Objective functions for optimization}

The suite of objective functions we use for goal-directed optimization were taken from \citeauthor{brown_guacamol:_2018}'s benchmarking function sets.\cite{brown_guacamol:_2018} Evaluation is divided into ``trivial'' tasks and ``hard'' tasks following the language of the original work. The trivial tasks are named as such because almost all molecular optimization methods can perform exceedingly well on them (thus they are not suitable for the assessment of generative models), whereas the hard tasks show greater variation as a function of the method used. However, all of these objective functions are relatively simple heuristic functions of molecular structure.

The trivial objectives we use include quantitative estimate of drug-likeness (QED);\cite{bickerton2012quantifying} a central nervous system (CNS) MPO\cite{wager2016central}; isomer of C$_{7}$H$_{8}$N$_{2}$O$_{2}$; and Pioglitazone MPO. The hard objectives we use include Osimertinib MPO, Fexofenadine MPO, Ranolazine MPO, Perindopril MPO, Amlodipine MPO, Ranolazine MPO , Sitagliptin MPO, Zaleplon MPO, Valsartan SMARTS, Scaffold Hop and Decorator Hop. Some MPO tasks try to identify molecules dissimilar to the titular molecule but with similar properties; other MPO tasks try to identify molecules similar to the titular molecule but with ``improved'' druglikeness properties. We didn't include the benchmarks that measure the similarity to commercial drug molecules and isomer benchmarks in hard tasks because we think they are less meaningful for drug discovery purposes. We refer readers to the list of benchmarks in ref.~\citenum{brown_guacamol:_2018} for a full description of these objectives.

\subsection{Biasing techniques for molecular generation}

\begin{itemize}

\item \textit{Post hoc} filtering is the approach where a CASP tool is used to filter unsynthesizable molecules suggested by an unbiased generation. We evaluate this approach by calculating the fraction of molecules that would pass the ASKCOS filter and their objective function values.

\item Training set biasing is the approach of starting with a molecule databases that has a higher fraction of synthesizable compounds as the training set for deep learning methods or the starting pool for genetic algorithms. In this paper, we use ChEMBL (68.3\% as tested) and MOSES (89.8\% as tested) as representative datasets with lower and higher synthesizabilities, respectively. This approach can be used in both unoptimized generation and optimized generation.

\item Heuristic biasing is the approach of modifying the main objective function to penalize the generation of unsynthesizable compounds. We apply a synthesizability function multiplier, ranging from 0 to 1, to a pre-normalized objective function (also ranging from 0 to 1). Specifically, we use a form of modified Gaussian and sigmoid function to rescale the heuristic score $x$:
\[
\textit{Modifier}=
\begin{cases}
1& \text{$x < \mu$}\\
e^{-\frac{(x - \mu)^2}{2 \sigma}}& \text{$x \geq \mu$}
\end{cases}
\]

\[
\textit{Modifier} = 1 - \frac{1}{1 + e^{a(x-b)}}
\]
We performed 30 iterations of Tree Parzen Estimator (TPE) Bayesian Optimization to determine the hyper-parameters for each score. The hyper-parameters aimed to maximize the fraction of synthesizable suggestions times the average of the objective function for the top 10 molecules from graph genetic algorithm. We tested the biasing effect of SA\_Score, SCScore, and length of SMILES string, but meaningful parameters could not be obtained for the SMILES string heuristic. The multipliers we use are shown in Figure \ref{fig:multipliers}. This approach can only be used in optimized generation.

\begin{itemize}
    \item SA\_Score\cite{ertl2009estimation} is a popular heuristic score for quantifying synthesizability. It computes a score using a fragment-contribution approach, where rarer fragments (as judged by their abundance in the PubChem database) are taken as an indication of lower synthesizability.
    \item SCScore\cite{coley2018scscore} is a learned synthetic complexity score computed by as neural network model trained on reaction data from the Reaxys database. It was designed with synthesis planning in mind to operate on molecules resembling not just drug-like products, but intermediates and simpler building blocks as well.
    \item SMILES length is a very simple heuristic that associates molecules with longer SMILES strings as an indication of synthetic difficulty. The length of a SMILES string correlates closely with the number of heavy atoms in a molecule (i.e., larger molecules are harder to synthesize), but is further increased by the presence of formal charges, ring closures, and defined stereochemistry.
\end{itemize}

\end{itemize}

\begin{acknowledgement}

This work was supported by the Machine Learning for Pharmaceutical Discovery and Synthesis consortium. We thank Mike Fortunato and Thomas Struble for assisting with programmatic interfacing of ASKCOS. We also thank Lagnajit Pattanaik and Klavs Jensen for commenting on the manuscript.

\end{acknowledgement}

\begin{suppinfo}

All code and data can be found at \url{https://github.com/wenhao-gao/askcos_synthesizability}. Additional results can be found in the supporting information.

\end{suppinfo}


\bibliography{paper}

%% graphical table of contents.
%%%%%%%%%%%%%%%%%%%%%%%%%%%%%%%%%%%%%%%%%%%%%%%%%%%%%%%%%%%%%%%%%%%%%

%% TODO
% \begin{tocentry}

% \centering
% \includegraphics[width=1.0\linewidth]{plots/tocentry4.png}

% \end{tocentry}

%%% AUTHOR ADDED
%%%%% REMOVE THE FOLLOWING LINES WHEN COMPILING FINAL VERSION
\newpage
\pagebreak
\renewcommand{\thefigure}{S\arabic{figure}}
\renewcommand{\thetable}{S\arabic{table}}
\setcounter{figure}{0} 
\setcounter{table}{0} 
\setcounter{page}{1}

\captionsetup{font=footnotesize}
\newgeometry{left=0.5in, right=0.5 in}

\begin{centering}

\textsf{\Large{\textbf{Supporting Information}}}

\vspace{0.5cm} 
\textsf{\Large{\textbf{The Synthesizability of Molecules Proposed by Generative Models}}}

\vspace{0.3cm}

% \textsf{\large{Wenhao Gao, Klavs F. Jensen$^\ast$ and Connor W. Coley$^\ast$}}
\textsf{\large{Wenhao Gao and Connor W. Coley$^\ast$}}
\begin{center}
{\it \small Department of Chemical Engineering, MIT, Cambridge, MA 02139 \\
\it \small Broad Institute of Harvard and MIT, Cambridge, MA 02139 \\
\it \small Department of Chemical and Biomolecular Engineering, Johns Hopkins University, Baltimore, MD 21218 
}  \\

% \textsf{\small E-mail: kfjensen@mit.edu; ccoley@mit.edu}
\textsf{\small E-mail: ccoley@mit.edu}
\end{center}

\end{centering}

\vspace{0.5cm}

\normalsize
\section{Code}

All code can be found at  \url{https://github.com/wenhao-gao/askcos_synthesizability}.

\section{Additional Results}

\subsection{Suitability of heuristic functions for estimating synthesizability}

\begin{figure}[h!]
    \centering
    \includegraphics[width=0.8\textwidth]{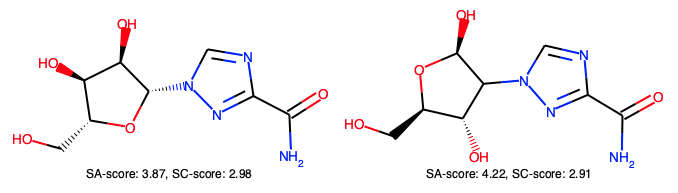}
    \caption{Illustration of the difficulty of applying heuristics to estimate synthesizability. Ribavirin (left) and its analogue (right) are structurally very similar, but their syntheses would be substantially different due to the inherent reactivity of ribose to favor substitution at the position leading to ribavirin. The SA\_Score and SCScore do not reflect that the righthand compound is much harder to access.}
    \label{fig:comp}
\end{figure}

\begin{figure}[h!]
    \centering
    \includegraphics[width=0.5\textwidth]{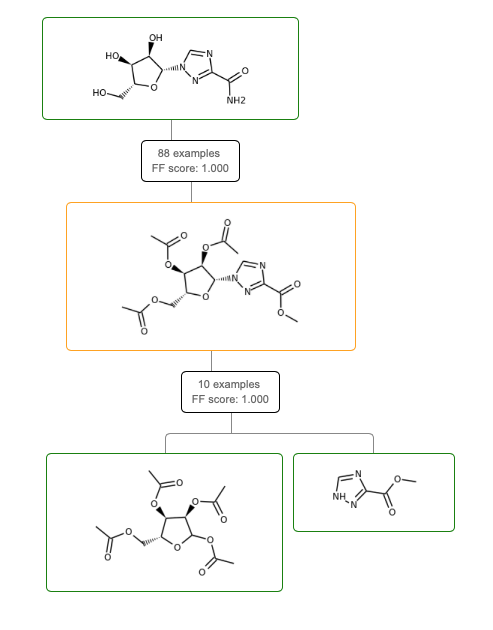}
    \caption{The synthetic pathway found by ASKCOS for ribavirin (left of Figure~\ref{fig:comp}). By explicitly planning  synthetic routes, ASKCOS easily distinguishes between the two compounds as it cannot identify a synthetic pathway for the righthand compound.}
    \label{fig:path}
\end{figure}

\begin{figure}[h!]
    \centering
    \includegraphics[width=0.6\textwidth]{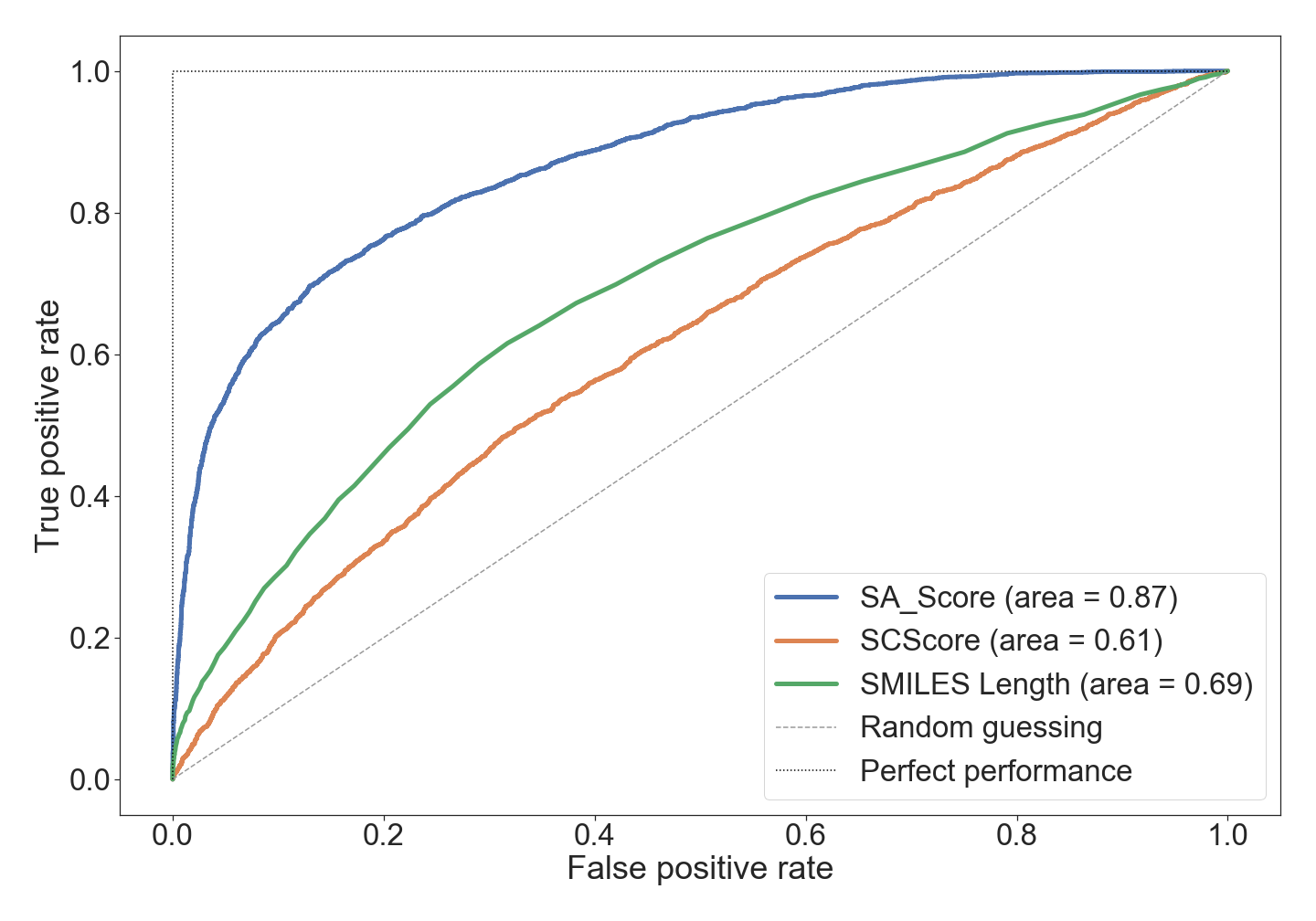}
    \caption{The receiver operating characteristic (ROC) curve obtained when using heuristic estimates of synthetic complexity for binary classification of molecules in the ``All but GDB'' compound set as synthesizable or unsynthesizable as perceived by ASKCOS. The area under the curve (AUC) quantifies the visual trends observed in Figure~\ref{fig:barplot}c-e. On this compound set, the SA\_Score outperforms the SMILES heuristic, which outperforms the SCScore. All three are better than randomly guessing.}
    \label{fig:roc}
\end{figure}

% \newpage
\FloatBarrier
\subsection{Full results of heuristics biasing in goal-directed generation}

\begin{figure}[h!]
    \centering
    \includegraphics[width=0.8\textwidth]{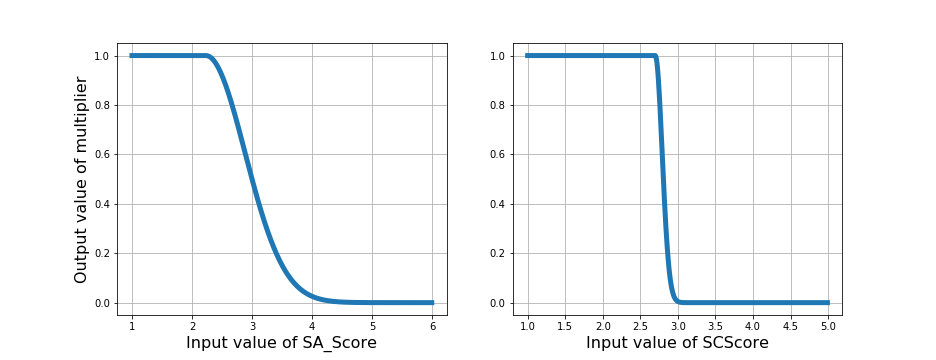}
    \caption{The scaled synthesizability multipliers used for heuristic biasing after optimizing shape parameters ($\mu$ and $\sigma$)  (see Methods).}
    \label{fig:multipliers}
\end{figure}

\begin{figure}[h!]
    \centering
    \includegraphics[width=0.9\textwidth]{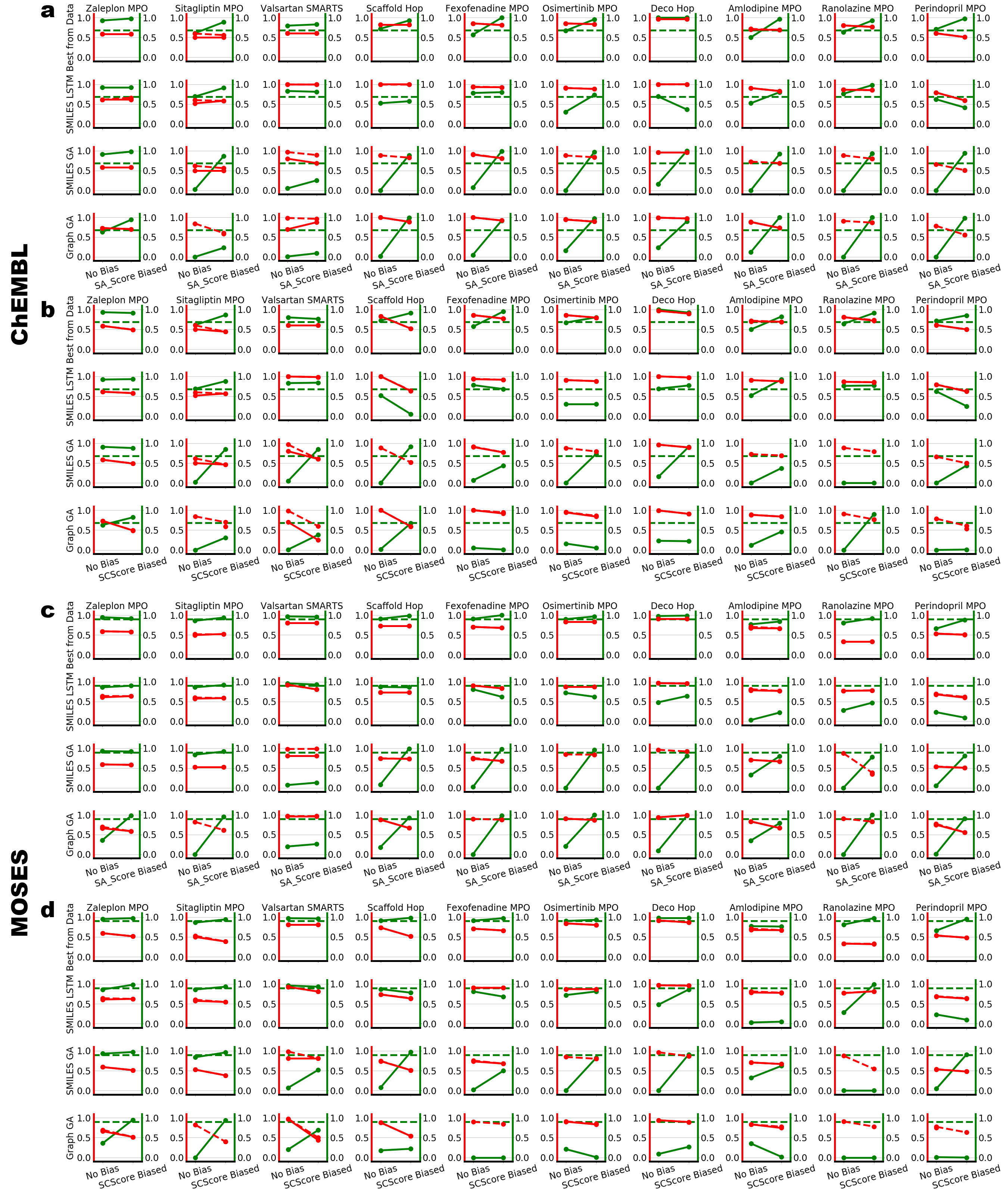}
    \caption{Change of synthesizability and objective function with heuristic biasing. (a) Biasing with SA\_Score after training on ChEMBL; (b) biasing with SCScore after training on ChEMBL; (c) biasing with SA\_Score after training on MOSES; (d) biasing with SCScore after training on MOSES. Within each panel, each row represents  one generative method; each column represents one objective function. In each plot, the green solid line represents the change of fraction of synthesizable compounds in the top-100, with the green dashed line as a reference for the synthesizability of the training set (ChEMBL or MOSES). Red solid lines represent the change in the objective function value of the top \emph{synthesizable} molecule, while the dashed red line represents the change in  objective function value of the top molecule, regardless of its synthesizability. Plots without a solid red line indicates that no synthesizable structure was obtained in the top 100 molecules. All plots have a dashed red line, though it might be occluded by the solid line.}
    \label{fig:plot_change}
\end{figure}

% \newpage

\begin{figure}[h!]
    \centering
    \includegraphics[width=\textwidth]{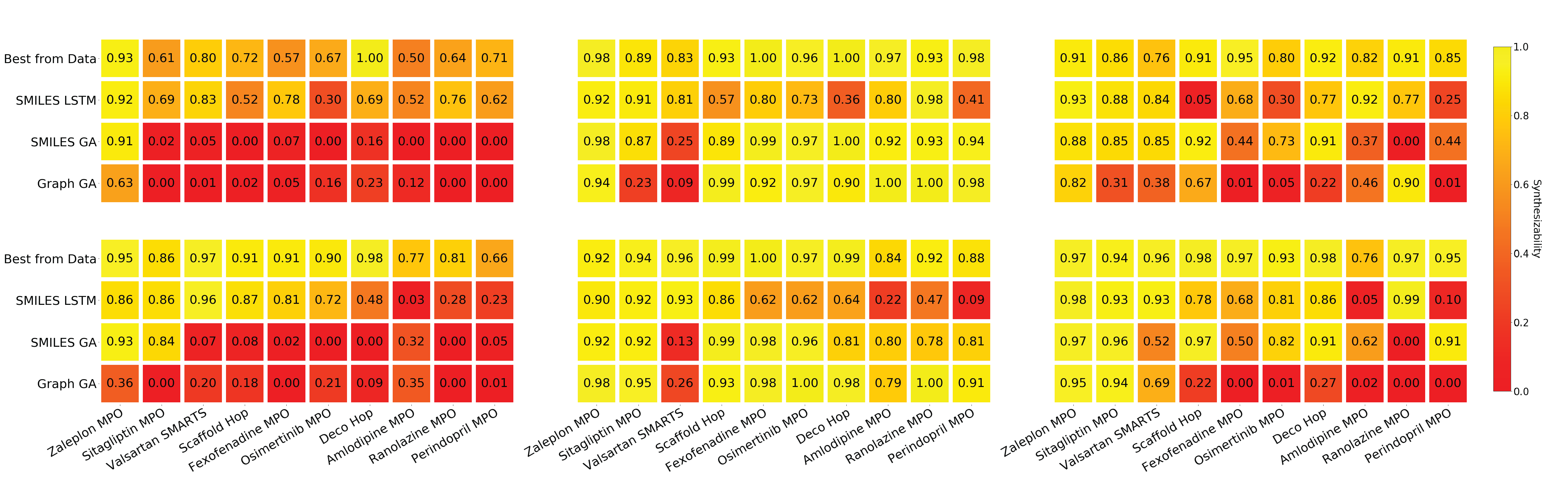}
    \caption{The synthesizable fraction of top-100 candidates proposed during goal-directed optimization for ``\textbf{hard}'' optimization tasks. (left) Without biasing; (middle) with heuristic biasing by SA\_Score; (right) with heuristic biasing by SCScore. (top) using ChEMBL for initial training; (botom) using MOSES for initial training. For many tasks, ASKCOS is unable to identify routes to a large fraction of generated molecules, particularly when using the Graph GA or SMILES GA methods.}
    \label{fig:hard_synth}
\end{figure}

\begin{figure}[h!]
    \centering
    \includegraphics[width=\textwidth]{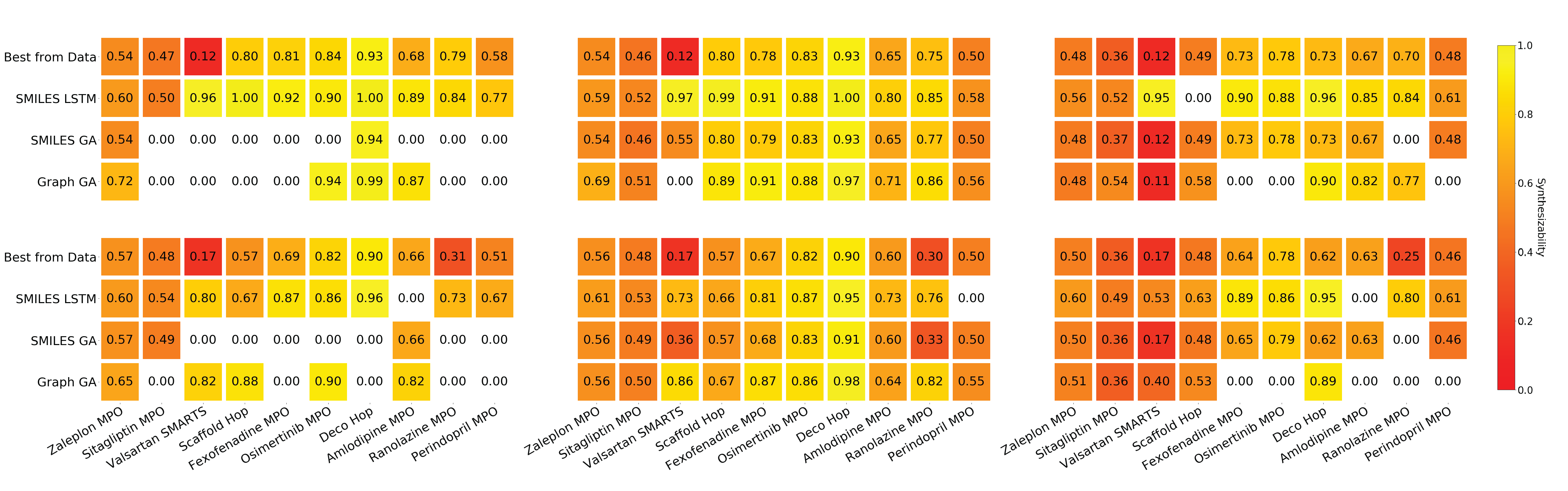}
    \caption{The average objective function value of the top-10 \emph{synthesizable} candidates, identified through \textit{post hoc} filtering of the top-100 candidates proposed during goal-directed optimization for ``\textbf{hard}'' optimization tasks. (left) Without biasing; (middle) with heuristic biasing by SA\_Score; (right) with heuristic biasing by SCScore. (top) using ChEMBL for initial training; (botom) using MOSES for initial training. White squares with a value of 0.00 indicates that fewer than 10 of the top 100 molecules were identified as synthesizable by ASKCOS. The top row showing the Best from Dataset represents a virtual screening approach.}
    \label{fig:hard_obj}
\end{figure}

% \newpage

\begin{figure}[h!]
    \centering
    \includegraphics[width=0.8\textwidth]{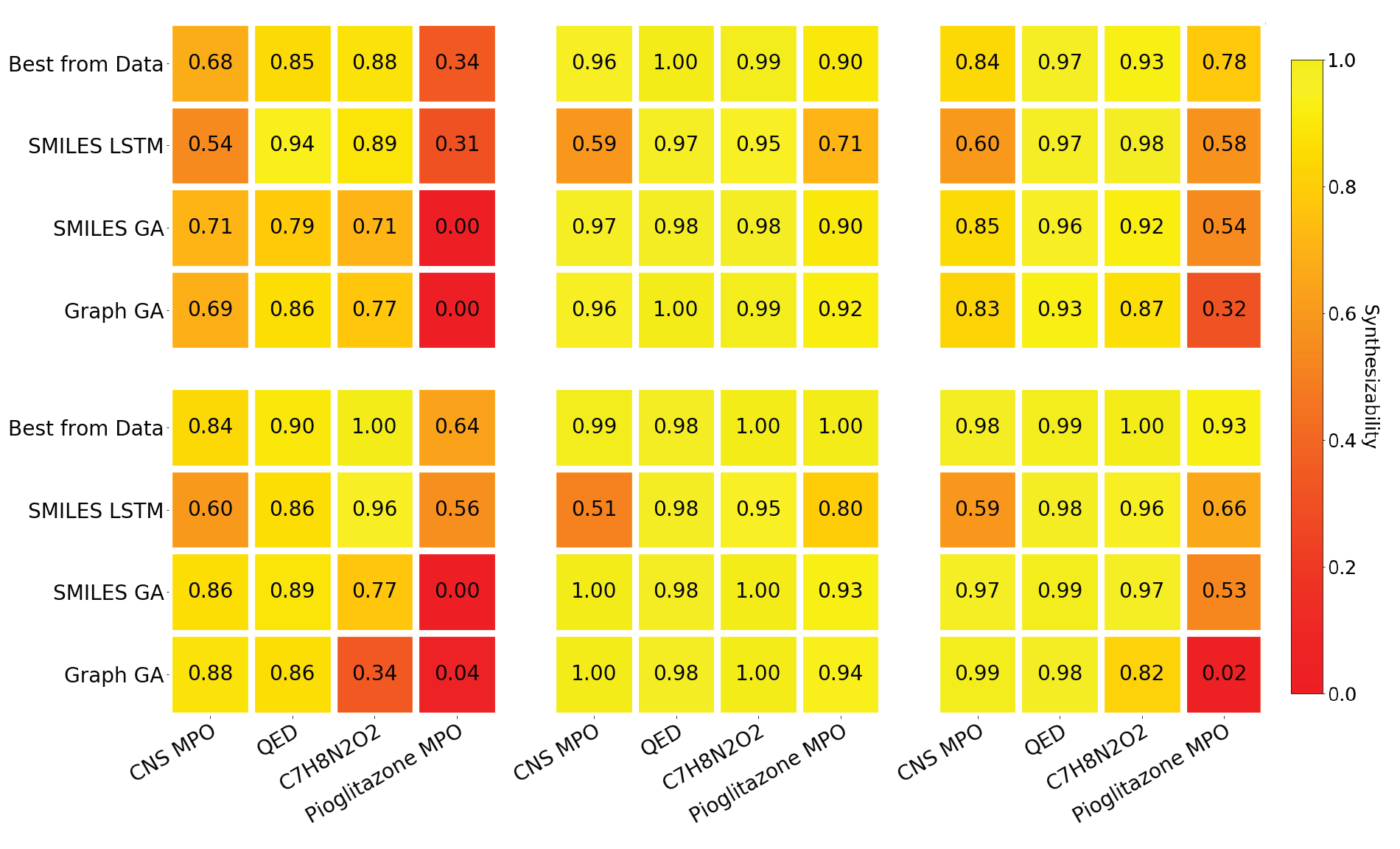}
    \caption{The synthesizable fraction of top-100 candidates proposed during goal-directed optimization for ``\textbf{trivial}'' optimization tasks. (left) Without biasing; (middle) with heuristic biasing by SA\_Score; (right) with heuristic biasing by SCScore. (top) using ChEMBL for initial training; (botom) using MOSES for initial training. Synthesizability is primarily an issue for the Pioglitazone MPO task.}
    \label{fig:trivial_synth}
\end{figure}

\begin{figure}[h!]
    \centering
    \includegraphics[width=0.8\textwidth]{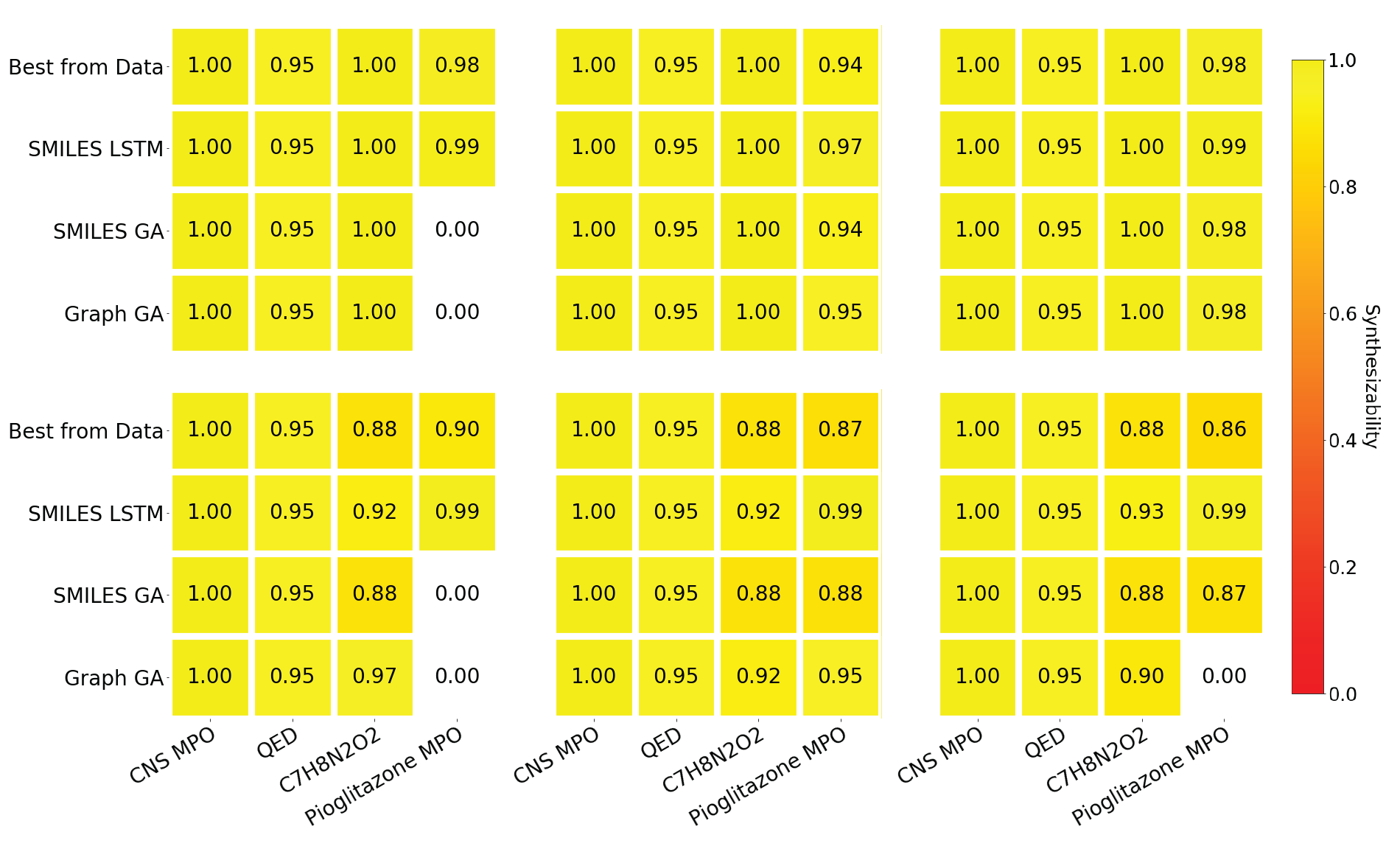}
    \caption{The average objective function value of the top-10 \emph{synthesizable} candidates, identified through \textit{post hoc} filtering of the top-100 candidates proposed during goal-directed optimization for ``\textbf{trivial}'' optimization tasks. (left) Without biasing; (middle) with heuristic biasing by SA\_Score; (right) with heuristic biasing by SCScore. (top) using ChEMBL for initial training; (botom) using MOSES for initial training. White squares with a value of 0.00 indicates that fewer than 10 of the top 100 molecules were identified as synthesizable by ASKCOS. The top row showing the Best from Dataset represents a virtual screening approach.}
    \label{fig:trivial_obj}
\end{figure}

% \newpage

\FloatBarrier
\subsection{Successful cases of heuristic biasing in goal-directed generation}

% \begin{figure}[h!]
%     \centering
%     \includegraphics[width=\textwidth]{Figure/nan2synth_1.png}
%     \caption{Molecules proposed during goal-directed optimization where there are no synthesizable structures proposed in the top 100 candidates in the absence of heuristic biasing. Each row represents a particular method, objective function, and initial training set. Columns show the best (unsynthesizable) molecule, the best synthesizable molecule after biasing with the SA\_Score, and the best synthesizable molecule after biasing with the SCScore. (a) SMILES GA / Ranolazine MPO / ChEMBL; (b) Graph GA / Ranolazine MPO / ChEMBL; (c) \placeholder; (d) \placeholder. Many of the structures proposed by the SMILES GA and Graph GA methods are nonsensical and clearly unsynthesizable, despite achieving a high objective function value. }
%     \label{fig:nan2synth_1}
% \end{figure}

\begin{figure}[h!]
\centering

\subfigure[SMILES GA / Ranolazine MPO / ChEMBL]{
\begin{minipage}[t]{0.9\linewidth}
\centering
\includegraphics[width=\textwidth]{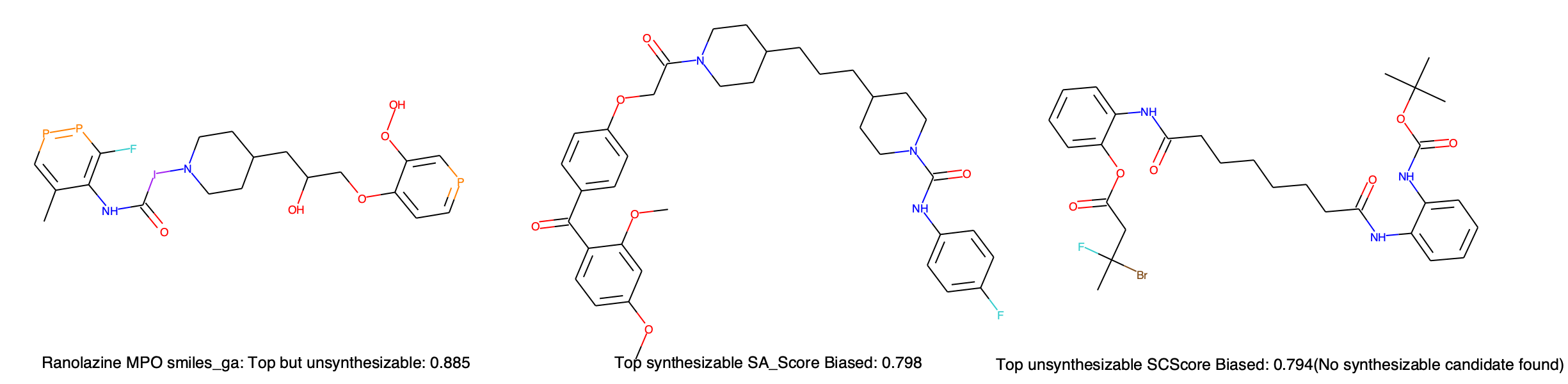}
%\caption{fig1}
\end{minipage}%
}

\subfigure[Graph GA / Ranolazine MPO / ChEMBL]{
\begin{minipage}[t]{0.9\linewidth}
\centering
\includegraphics[width=\textwidth]{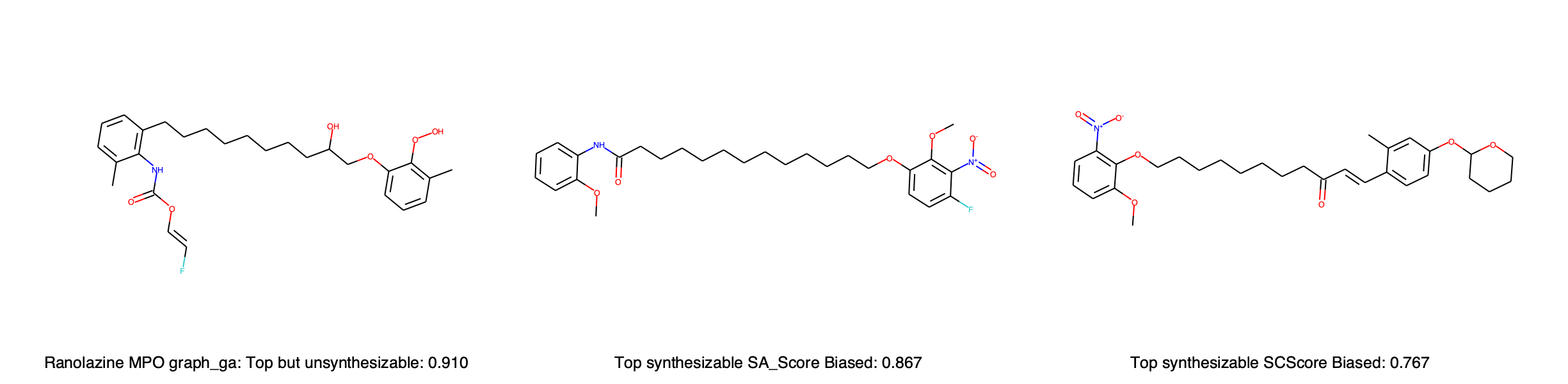}
%\caption{fig2}
\end{minipage}%
}

\subfigure[SMILES GA / Perindopril MPO / ChEMBL]{
\begin{minipage}[t]{0.9\linewidth}
\centering
\includegraphics[width=\textwidth]{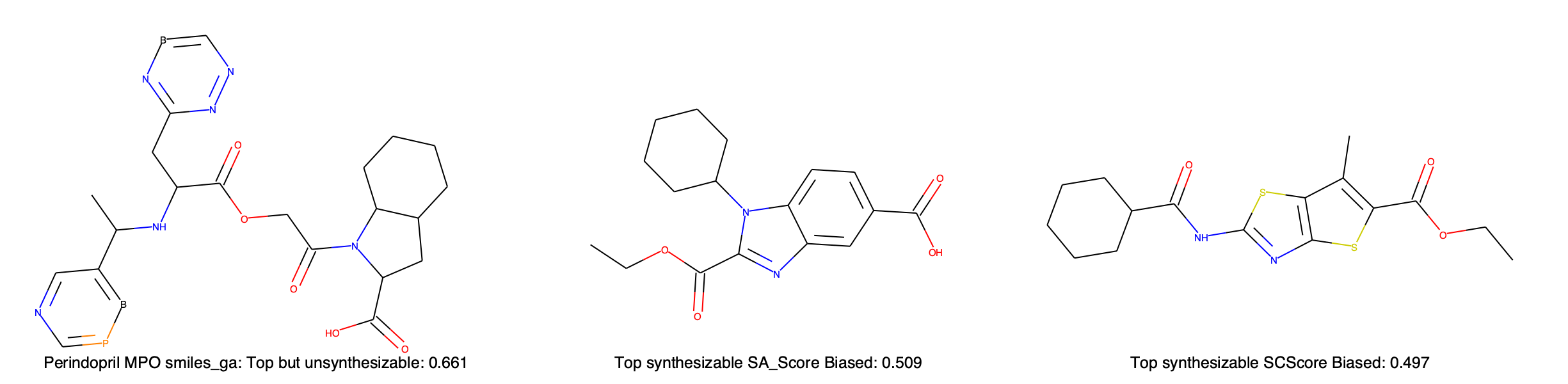}
%\caption{fig2}
\end{minipage}%
}

\subfigure[Graph GA / Perindopril MPO / ChEMBL]{
\begin{minipage}[t]{0.9\linewidth}
\centering
\includegraphics[width=\textwidth]{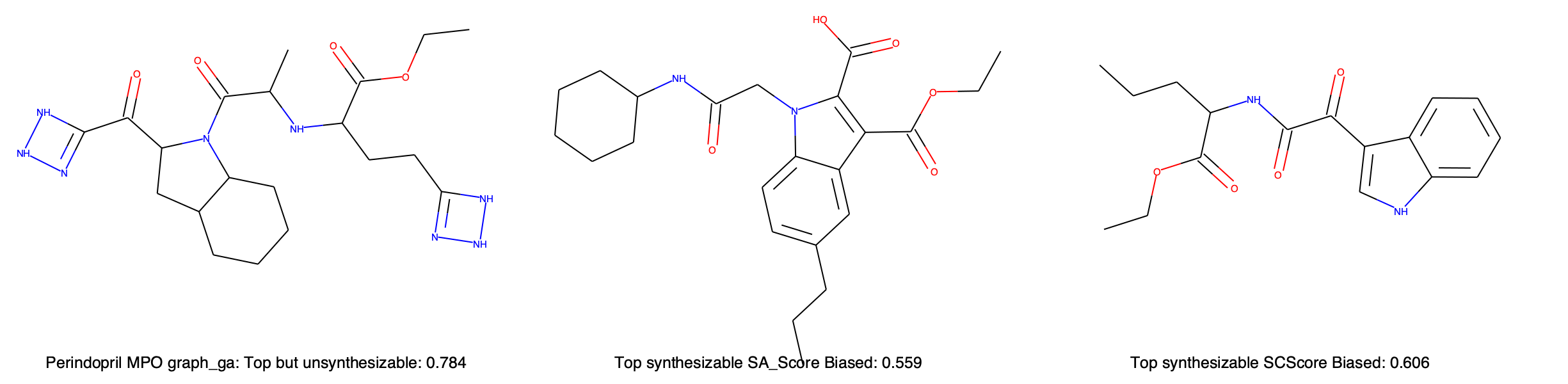}
%\caption{fig2}
\end{minipage}%
}

\centering
\caption{Molecules proposed during goal-directed optimization where there are no synthesizable structures proposed in the top 100 candidates in the absence of heuristic biasing. Each row represents a particular method, objective function, and initial training set. From left to right, we draw the best (unsynthesizable) molecule, the best synthesizable molecule after biasing with the SA\_Score, and the best synthesizable molecule after biasing with the SCScore. Many of the structures proposed by the SMILES GA and Graph GA methods are nonsensical and clearly unsynthesizable, despite achieving a high objective function value.}
\label{fig:nan2synth_1}
\end{figure}

\begin{figure}[h!]
\centering

\subfigure[SMILES GA / Amlodipine MPO / ChEMBL]{
\begin{minipage}[t]{0.9\linewidth}
\centering
\includegraphics[width=\textwidth]{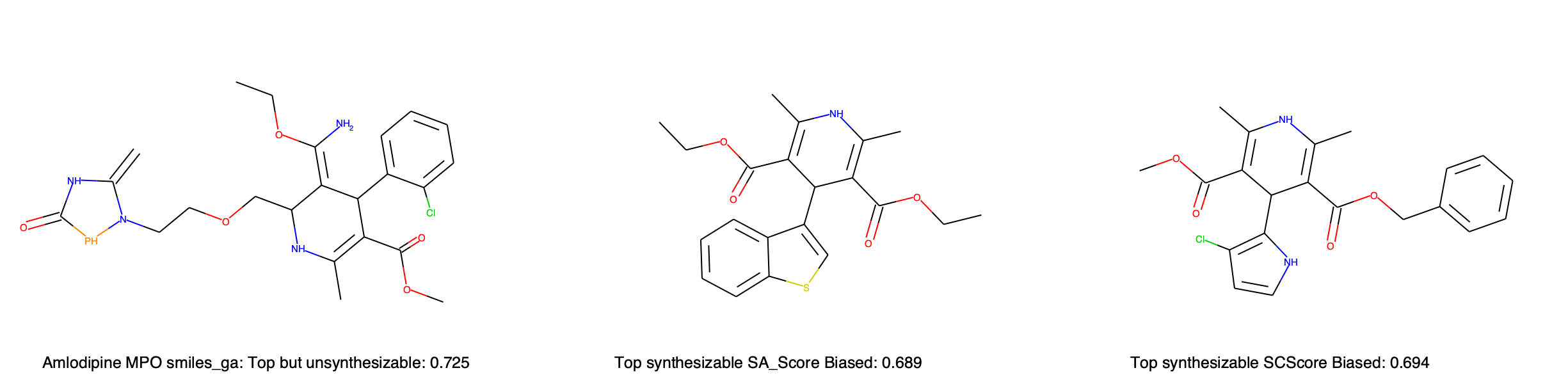}
%\caption{fig1}
\end{minipage}%
}

\subfigure[Graph GA / Sitagliptin MPO / ChEMBL]{
\begin{minipage}[t]{0.9\linewidth}
\centering
\includegraphics[width=\textwidth]{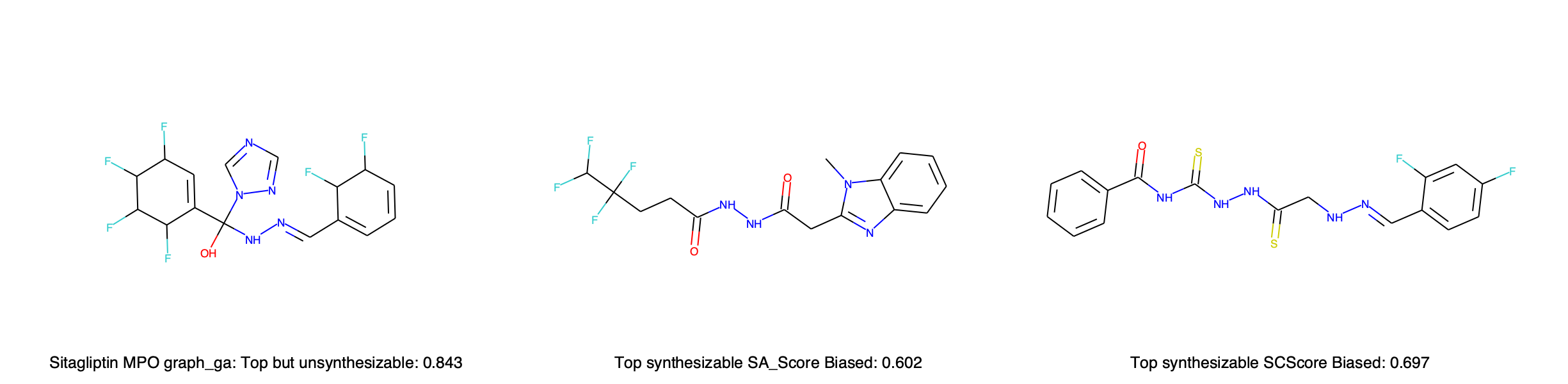}
%\caption{fig2}
\end{minipage}%
}

\subfigure[SMILES GA / Scaffold Hop / ChEMBL]{
\begin{minipage}[t]{0.9\linewidth}
\centering
\includegraphics[width=\textwidth]{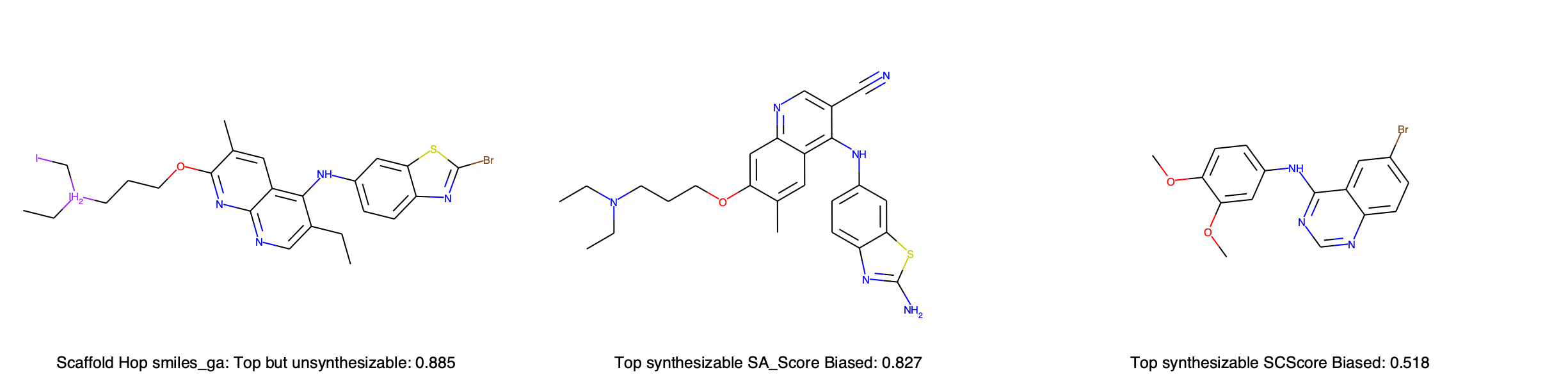}
%\caption{fig2}
\end{minipage}%
}

\subfigure[SMILES GA / Osimertinib MPO / MOSES]{
\begin{minipage}[t]{0.9\linewidth}
\centering
\includegraphics[width=\textwidth]{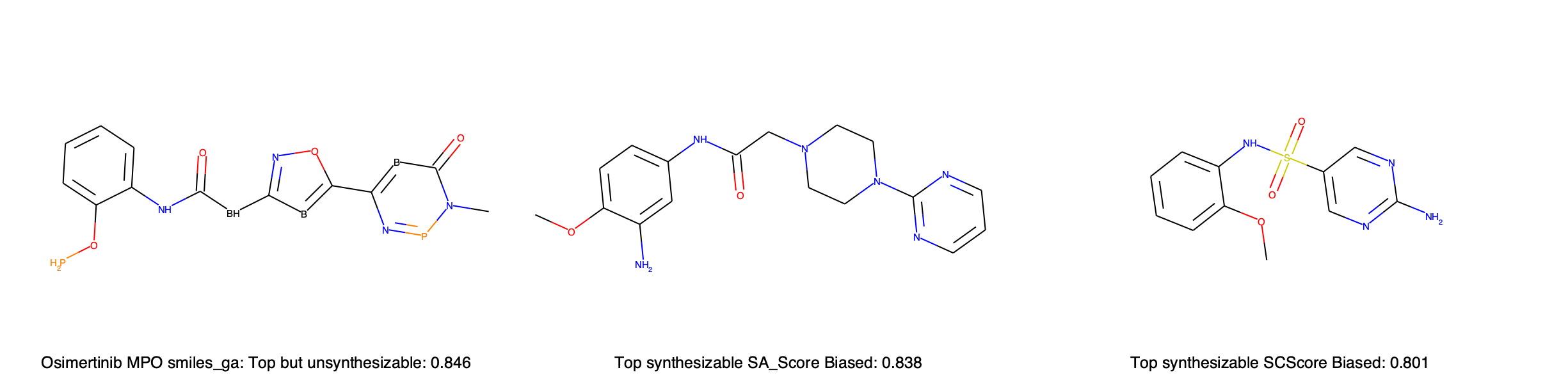}
%\caption{fig2}
\end{minipage}%
}

\centering
\caption{Molecules proposed during goal-directed optimization where there are no synthesizable structures proposed in the top 100 candidates in the absence of heuristic biasing. Each row represents a particular method, objective function, and initial training set. From left to right, we draw the best (unsynthesizable) molecule, the best synthesizable molecule after biasing with the SA\_Score, and the best synthesizable molecule after biasing with the SCScore. Many of the structures proposed by the SMILES GA and Graph GA methods are nonsensical and clearly unsynthesizable, despite achieving a high objective function value.}
\label{fig:nan2synth_2}
\end{figure}

\begin{figure}[h!]
\centering

\subfigure[Graph GA / Fexofenadine MPO / MOSES]{
\begin{minipage}[t]{0.9\linewidth}
\centering
\includegraphics[width=\textwidth]{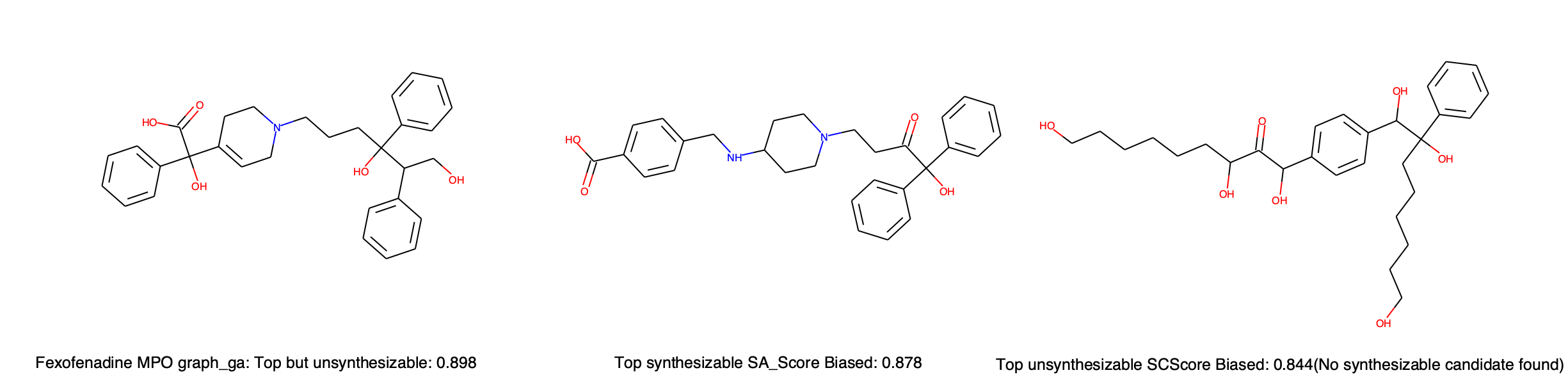}
%\caption{fig1}
\end{minipage}%
}

\subfigure[SMILES GA / Ranolazine MPO / MOSES]{
\begin{minipage}[t]{0.9\linewidth}
\centering
\includegraphics[width=\textwidth]{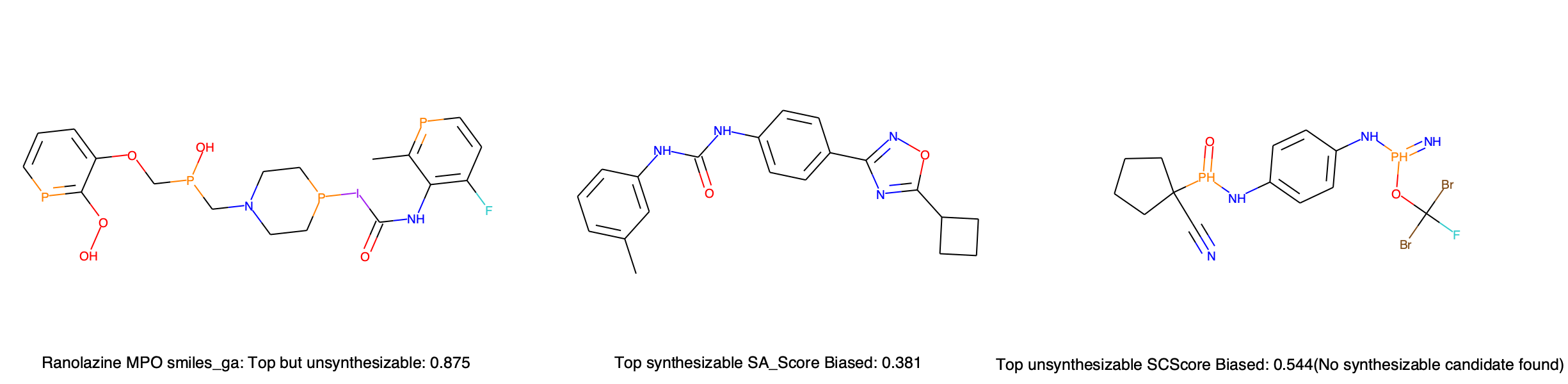}
%\caption{fig2}
\end{minipage}%
}

\subfigure[Graph GA / Ranolazine MPO / MOSES]{
\begin{minipage}[t]{0.9\linewidth}
\centering
\includegraphics[width=\textwidth]{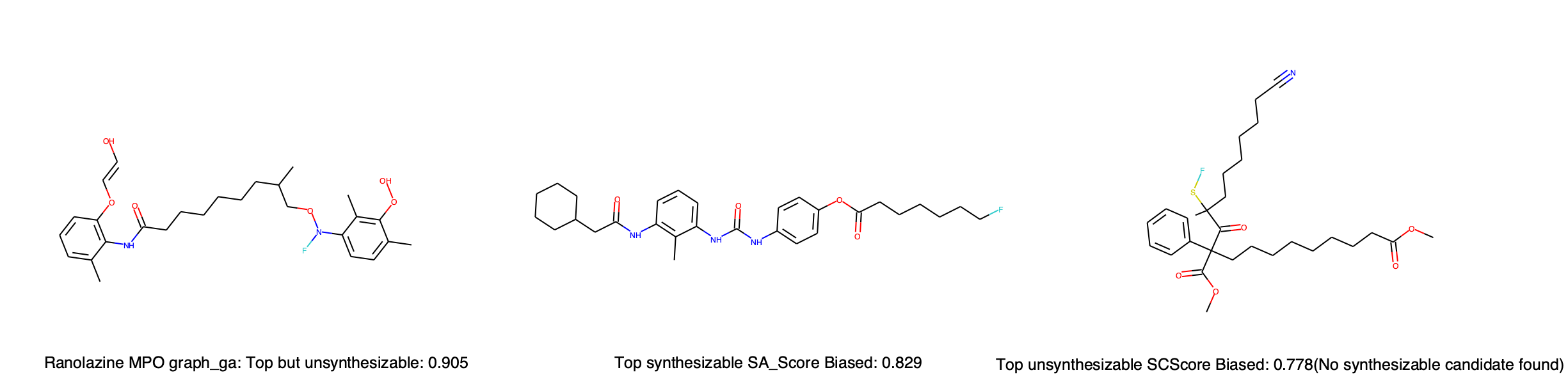}
%\caption{fig2}
\end{minipage}%
}

\subfigure[SMILES GA / Decoration Hop / MOSES]{
\begin{minipage}[t]{0.9\linewidth}
\centering
\includegraphics[width=\textwidth]{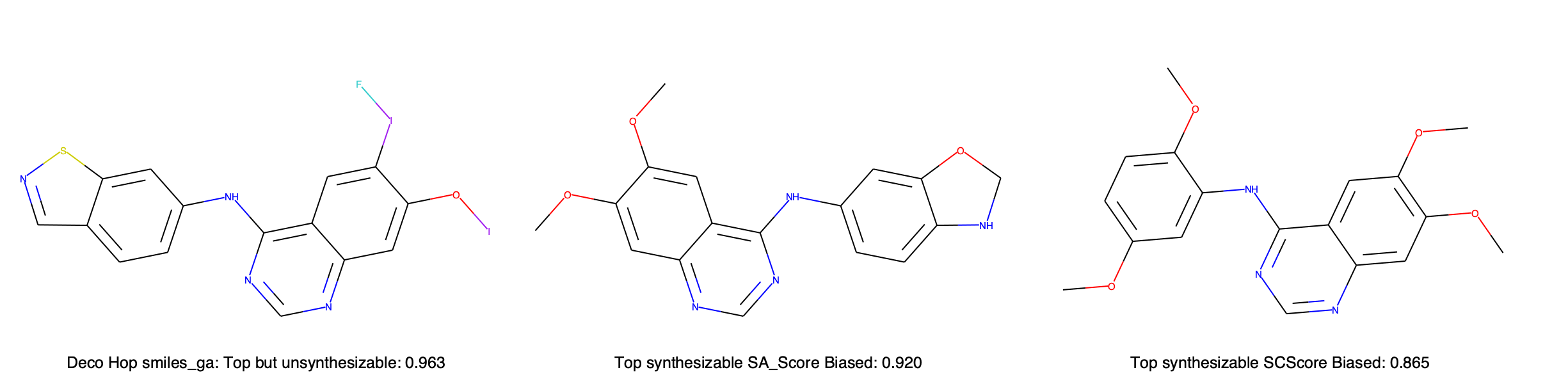}
%\caption{fig2}
\end{minipage}%
}

\centering
\caption{Molecules proposed during goal-directed optimization where there are no synthesizable structures proposed in the top 100 candidates in the absence of heuristic biasing. Each row represents a particular method, objective function, and initial training set. From left to right, we draw the best (unsynthesizable) molecule, the best synthesizable molecule after biasing with the SA\_Score, and the best synthesizable molecule after biasing with the SCScore.}
\label{fig:nan2synth_3}
\end{figure}

\begin{figure}[h!]
\centering

\subfigure[SMILES LSTM / Sitagliptin MPO / ChEMBL]{
\begin{minipage}[t]{\linewidth}
\centering
\includegraphics[width=\textwidth]{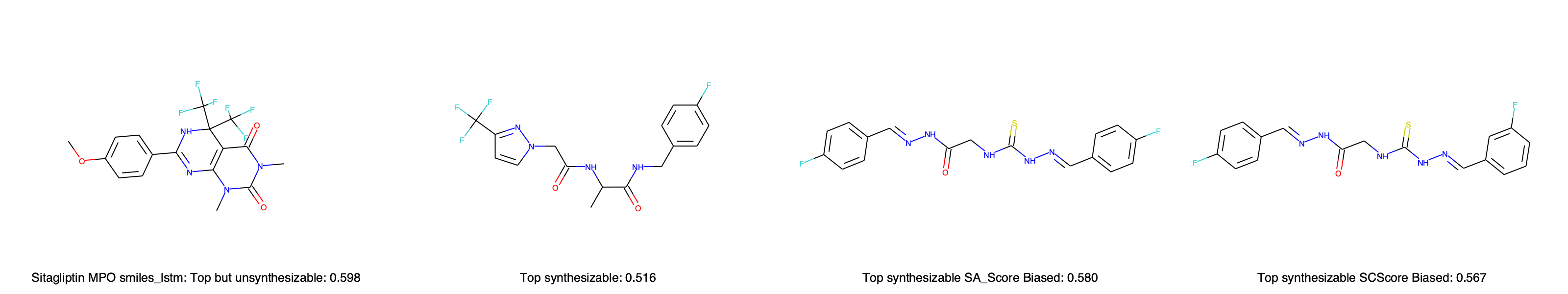}
%\caption{fig1}
\end{minipage}%
}

\subfigure[SMILES LSTM / Ranolazine MPO / MOSES]{
\begin{minipage}[t]{\linewidth}
\centering
\includegraphics[width=\textwidth]{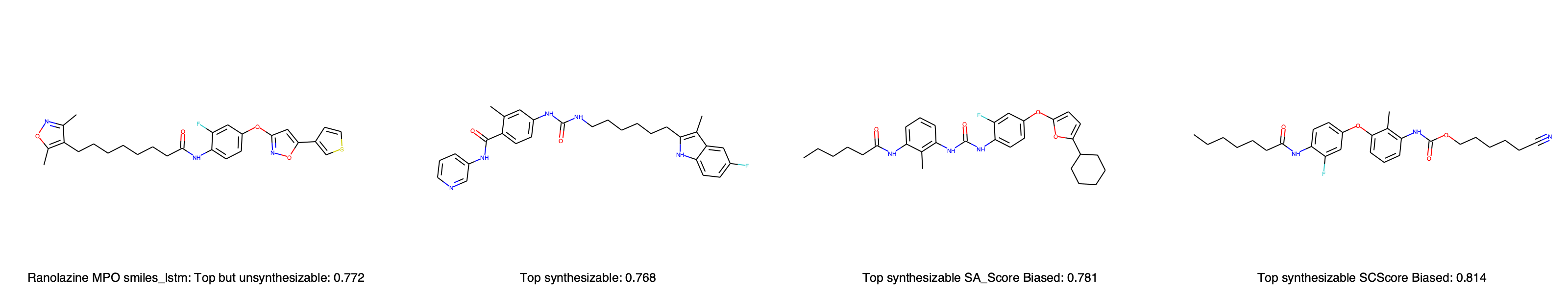}
%\caption{fig2}
\end{minipage}%
}

\subfigure[SMILES LSTM / Sitagliptin MPO / MOSES]{
\begin{minipage}[t]{\linewidth}
\centering
\includegraphics[width=\textwidth]{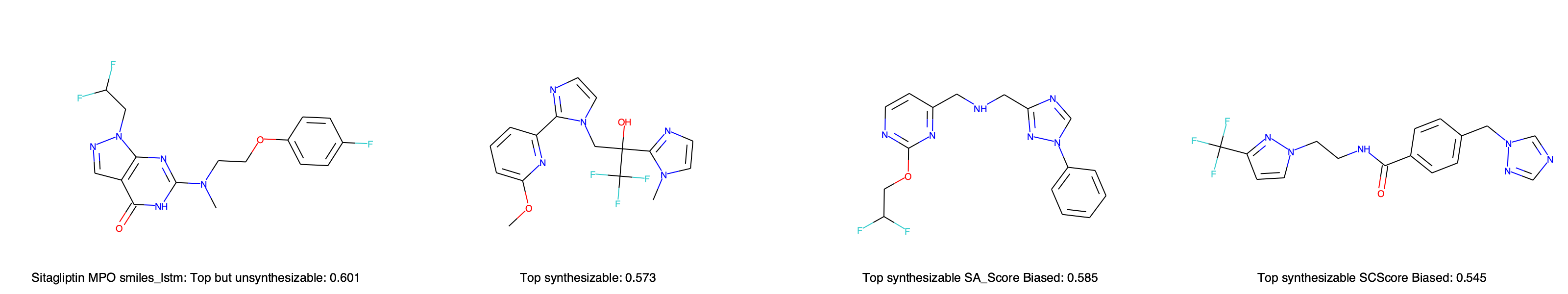}
%\caption{fig2}
\end{minipage}%
}

\subfigure[SMILES LSTM / Zaleplon MPO / MOSES]{
\begin{minipage}[t]{\linewidth}
\centering
\includegraphics[width=\textwidth]{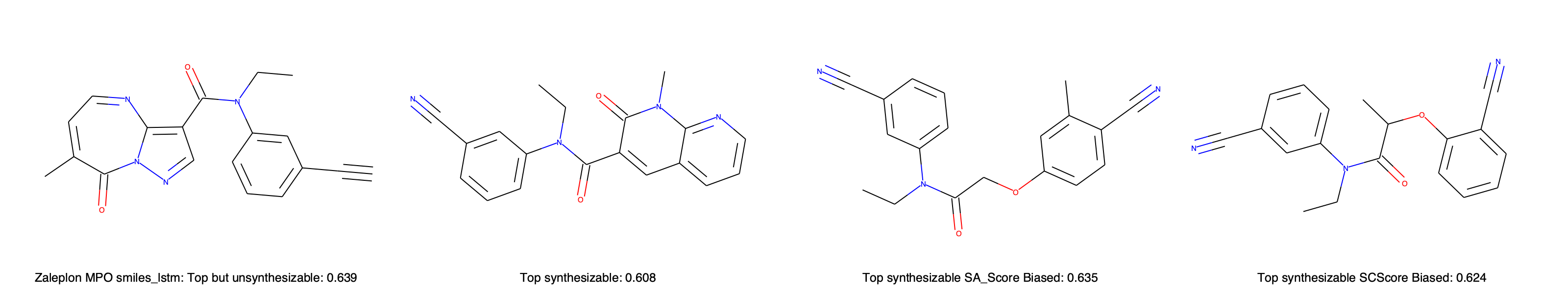}
%\caption{fig2}
\end{minipage}%
}

\centering
\caption{Molecules proposed during goal-directed optimization where the main objective function value of the top-1 \emph{synthesizable} structure is improved by heuristic biasing. Each row represents a particular method, objective function, and initial training set. From left to right, we draw the best (unsynthesizable) molecule, the best synthesizable molecule, the best synthesizable molecule after biasing with the SA\_Score, and the best synthesizable molecule after biasing with the SCScore. All cases shown here use the SMILES LSTM method.}
\label{fig:synth2improve}
\end{figure}

% \begin{figure}[h!]
%     \centering
%     \includegraphics[width=\textwidth]{Figure/nan2synth_2.png}
%     \caption{Same as \ref{fig:nan2synth_1}}
%     \label{fig:nan2synth_2}
% \end{figure}

% \begin{figure}[h!]
%     \centering
%     \includegraphics[width=\textwidth]{Figure/nan2synth_3.png}
%     \caption{Same as \ref{fig:nan2synth_1}}
%     \label{fig:nan2synth_3}
% \end{figure}

% \begin{figure}[h!]
%     \centering
%     \includegraphics[width=\textwidth]{Figure/synth2improve_cm.png}
%     \caption{Cases that the objective function of top-1 synthesizable structure was improved by biasing.}
%     \label{fig:synth2improve}
% \end{figure}

% \newpage
\FloatBarrier
\subsection{Agreement between ASKCOS results and synthesizability heuristics}

\begin{figure}[h!]
    \centering
    \includegraphics[width=0.7\textwidth]{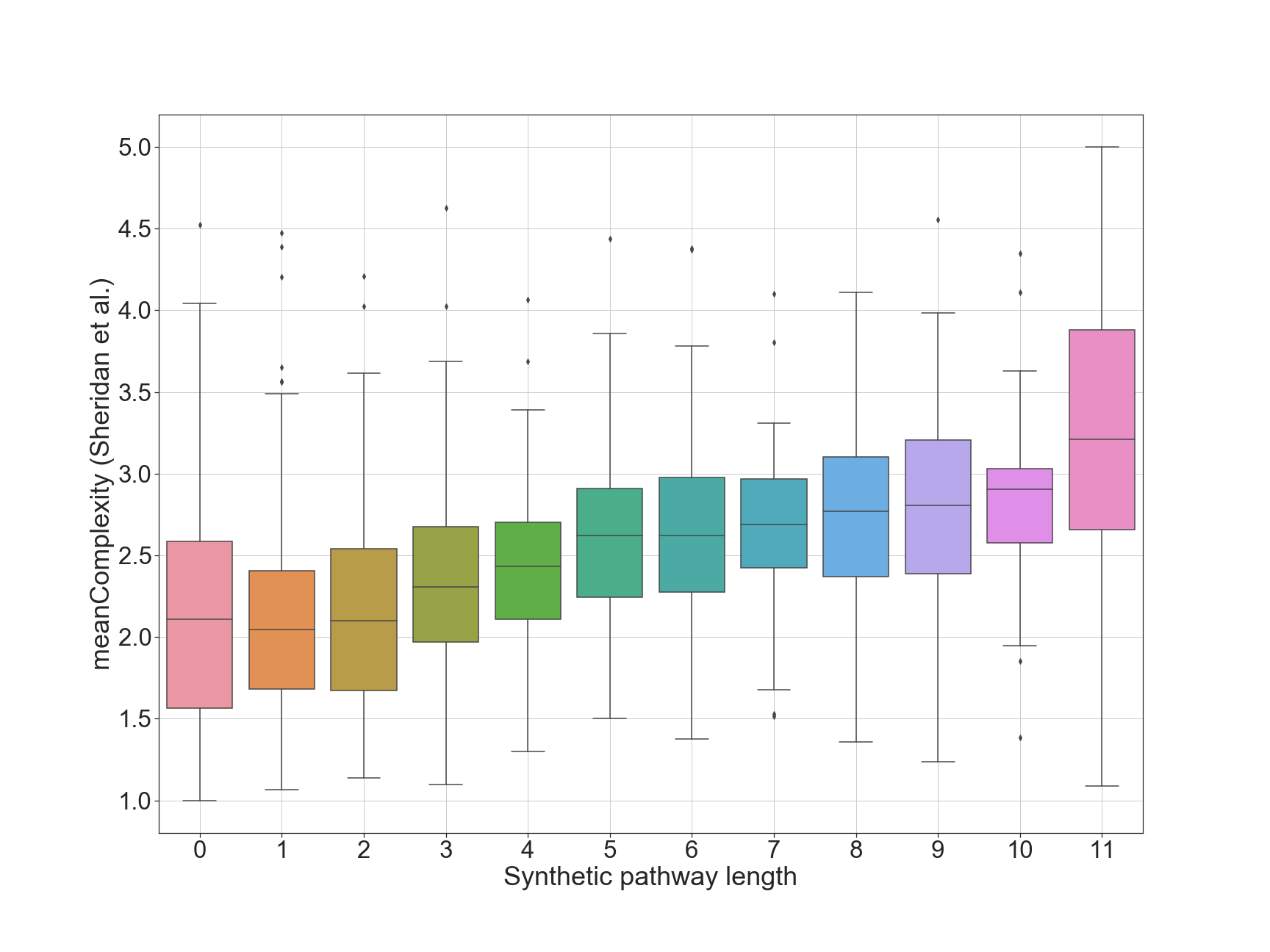}
    \caption{Correlation between the length of the first synthetic pathway found by ASKCOS and expert scores assigned by chemists in \citeauthor{sheridan_modeling_2014}\cite{sheridan_modeling_2014}. A length of 0 indicates that the molecule can be found in our database of readily-purchasable compounds; a length of 11 indicates that no pathway was found with the fixed expansion settings (see Methods).}
    \label{fig:depth_mc}
\end{figure}

\begin{figure}[h!]
    \centering
    \includegraphics[width=0.7\textwidth]{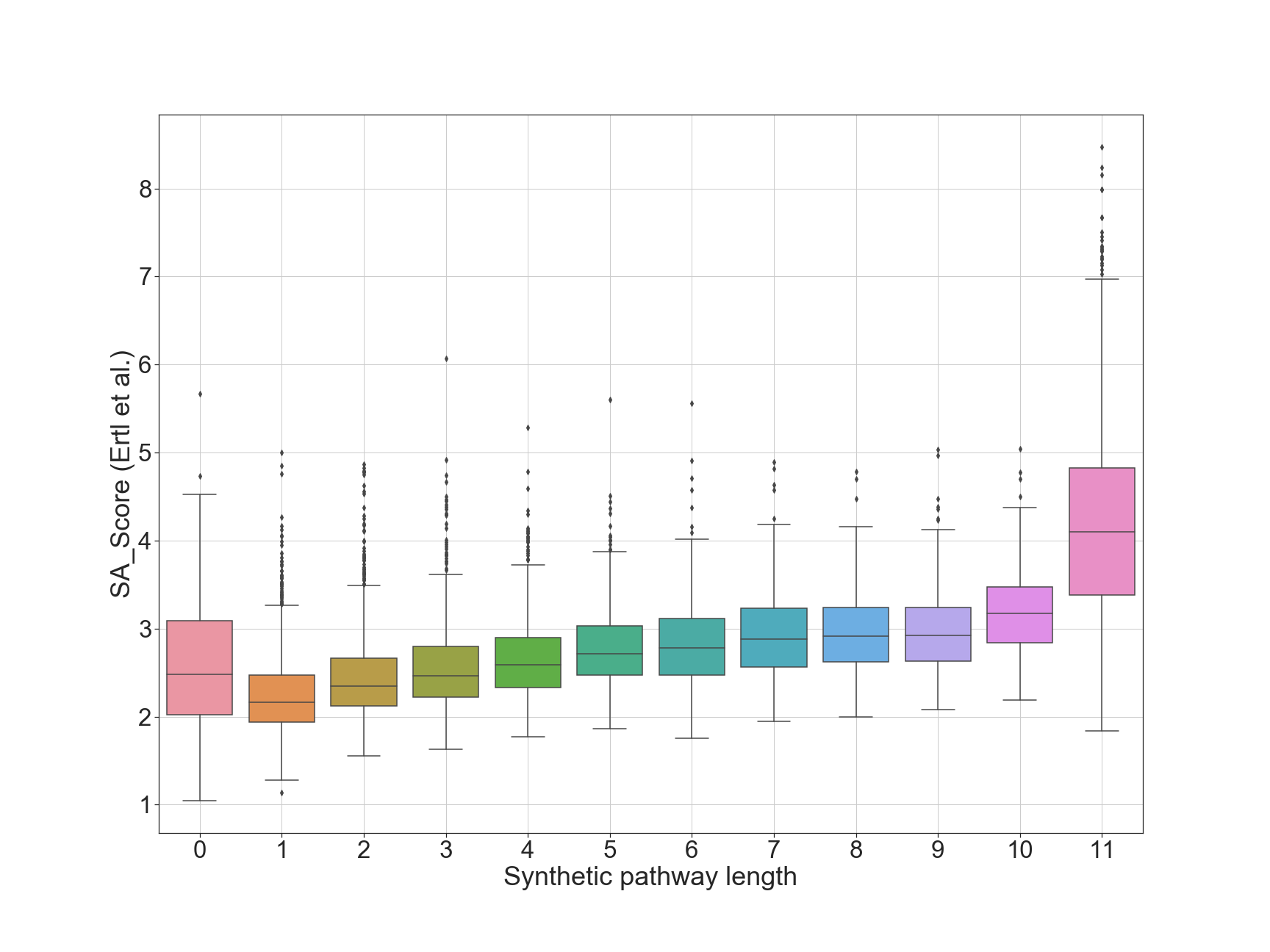}
    \caption{Correlation between the length of the first synthetic pathway found by ASKCOS using all compound datasets and the SA\_Score\cite{ertl2009estimation} heuristic. A length of 0 indicates that the molecule can be found in our database of readily-purchasable compounds; a length of 11 indicates that no pathway was found with the fixed expansion settings (see Methods).}
    \label{fig:depth_sa}
\end{figure}

\begin{figure}[h!]
    \centering
    \includegraphics[width=0.7\textwidth]{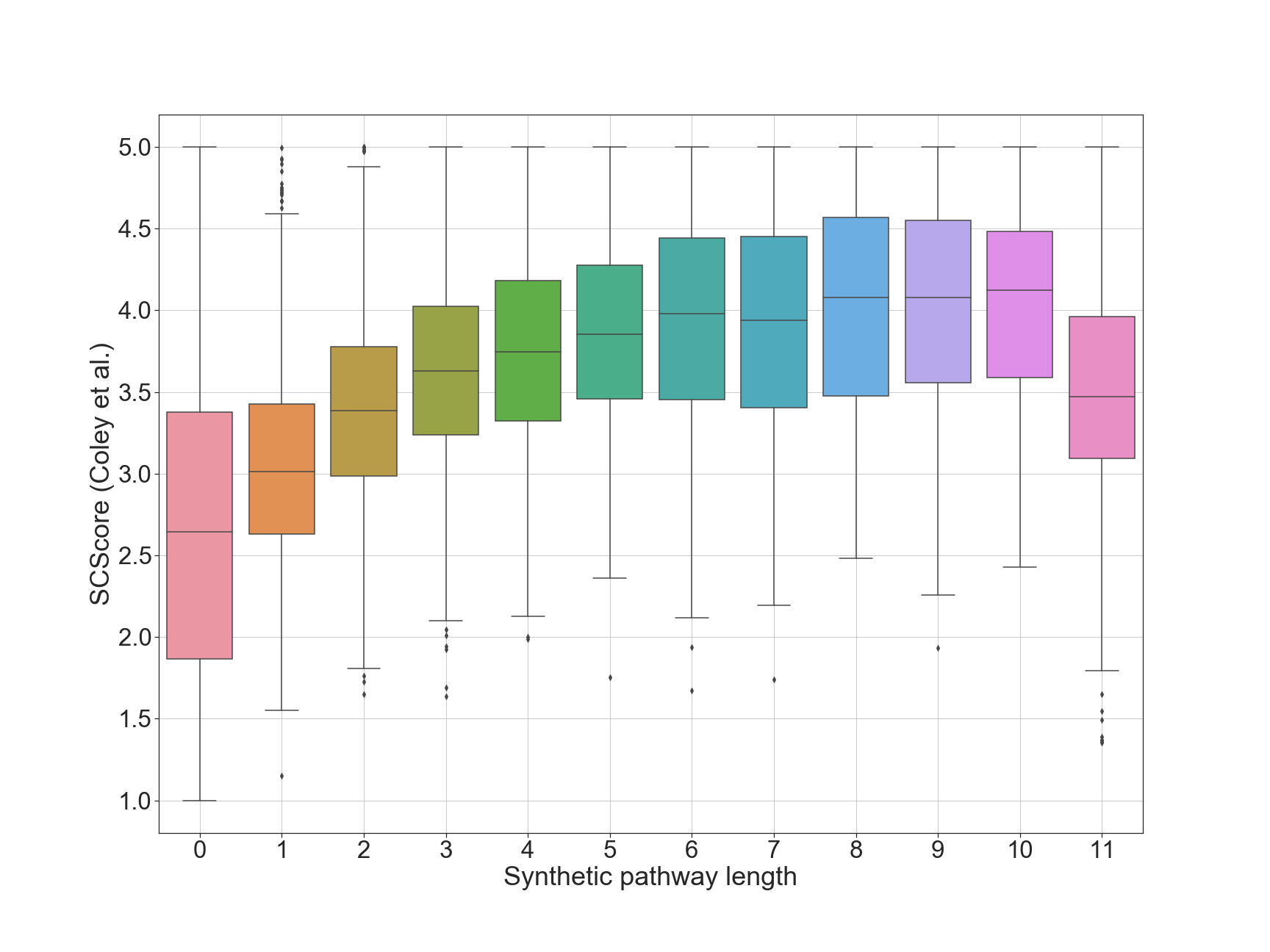}
    \caption{Correlation between the length of the first synthetic pathway found by ASKCOS using all compound datasets and the SCScore\cite{coley2018scscore} heuristic A length of 0 indicates that the molecule can be found in our database of readily-purchasable compounds; a length of 11 indicates that no pathway was found with the fixed expansion settings (see Methods).}
    \label{fig:depth_sc}
\end{figure}

\begin{figure}[h!]
    \centering
    \includegraphics[width=0.7\textwidth]{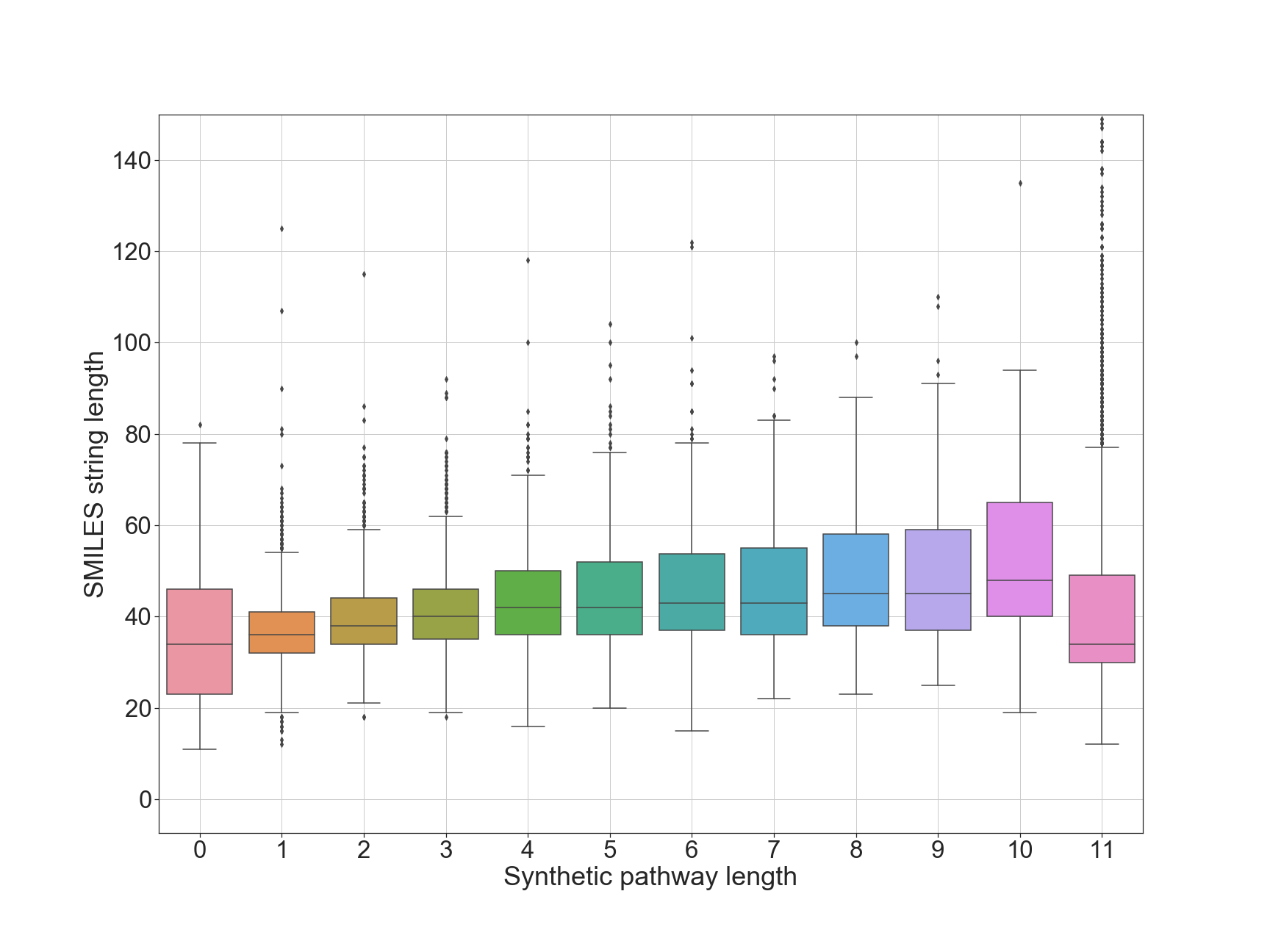}
    \caption{Correlation between the length of the first synthetic pathway found by ASKCOS using all compound datasets and the SMILES length heuristic A length of 0 indicates that the molecule can be found in our database of readily-purchasable compounds; a length of 11 indicates that no pathway was found with the fixed expansion settings (see Methods).}
    \label{fig:depth_smi}
\end{figure}

\begin{figure}[h!]
    \centering
    \includegraphics[width=0.5\textwidth]{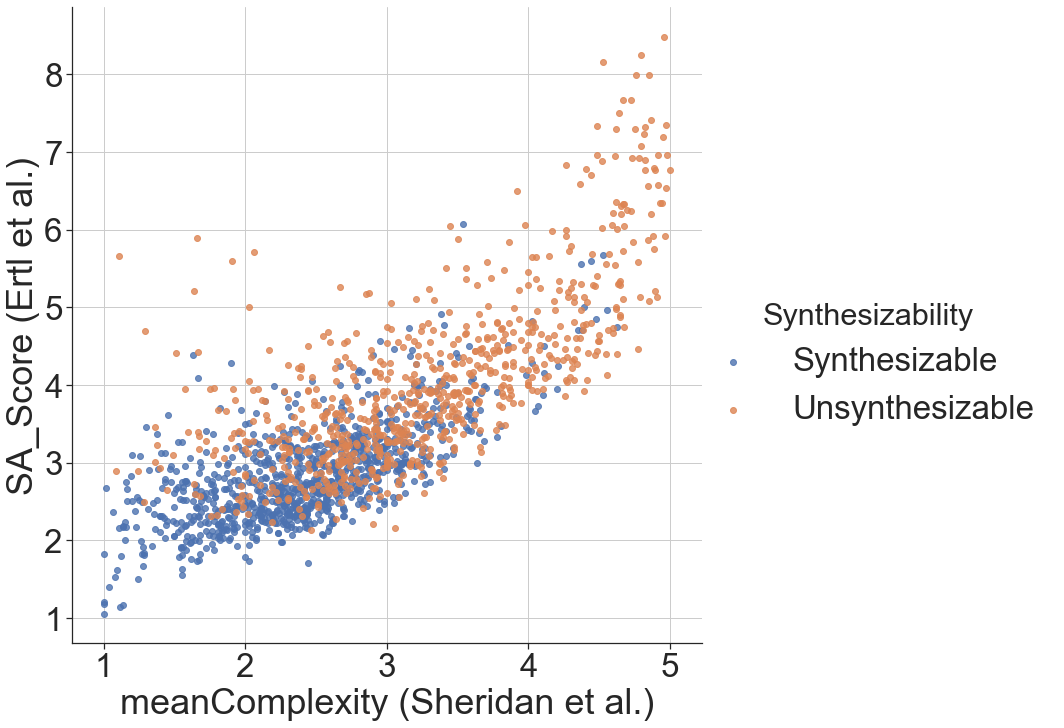}
    \caption{Correlation between \citeauthor{sheridan_modeling_2014}'s meanComplexity\cite{sheridan_modeling_2014} and the SA\_Score\cite{ertl2009estimation}; each compound is colored by its synthesizability according to ASKCOS.}
    \label{fig:sa_mc}
\end{figure}

\begin{figure}[h!]
    \centering
    \includegraphics[width=0.5\textwidth]{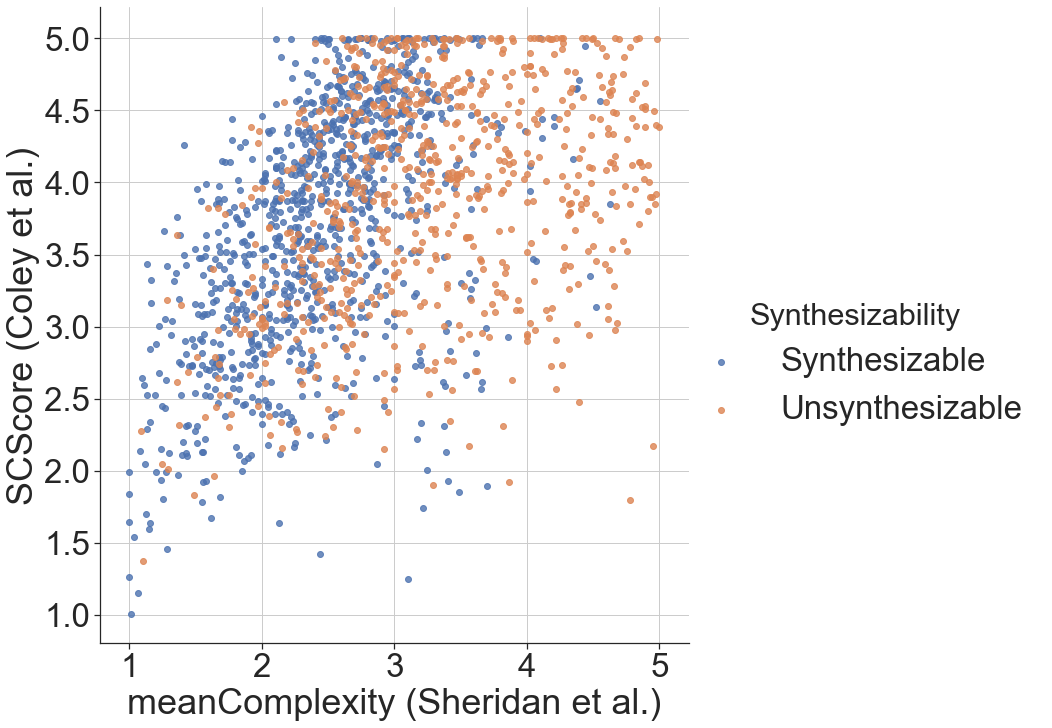}
    \caption{Correlation between \citeauthor{sheridan_modeling_2014}'s meanComplexity\cite{sheridan_modeling_2014} and the SCScore\cite{coley2018scscore}; each compound is colored by its synthesizability according to ASKCOS.}
    \label{fig:sc_mc}
\end{figure}

\begin{figure}[h!]
    \centering
    \includegraphics[width=0.5\textwidth]{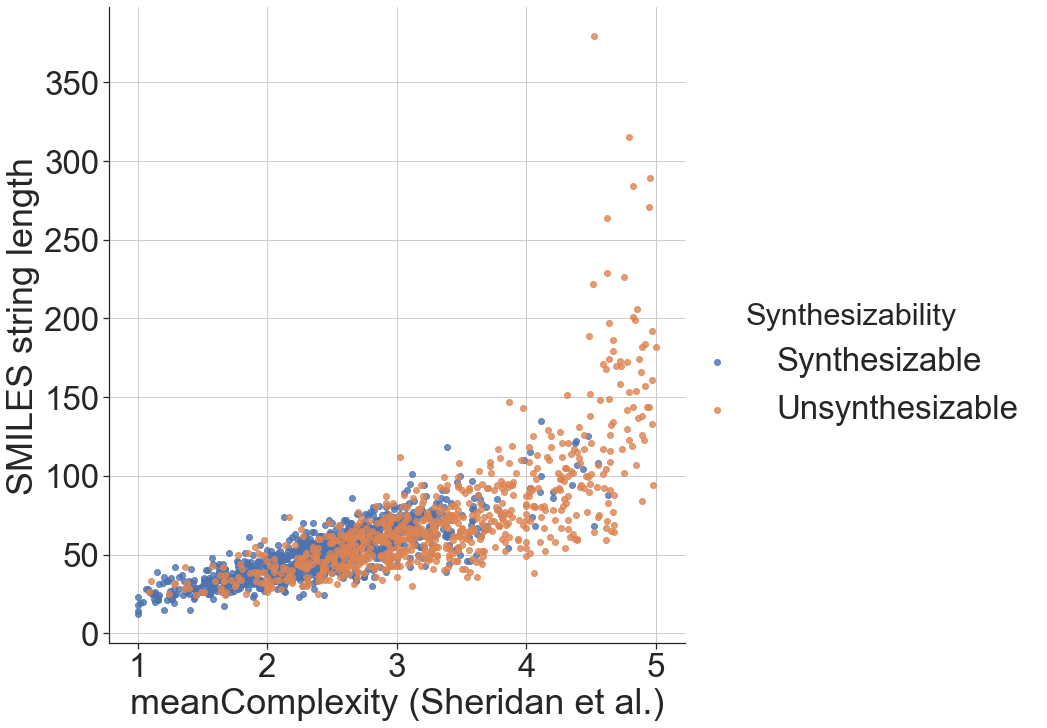}
    \caption{Correlation between \citeauthor{sheridan_modeling_2014}'s meanComplexity\cite{sheridan_modeling_2014} and the SMILES string length; each compound is colored by its synthesizability according to ASKCOS.}
    \label{fig:smi_mc}
\end{figure}

% Figures showing pathways to the optimized molecules for different goal-directed benchmarks 

\end{document}